# Augmenting Zero Trust Architecture to Endpoints Using Blockchain: A State-of-The-Art Review

Lampis Alevizos
*School of Psychology and Computer Science, University of Central Lancashire (UCLan)*
Preston, UK
lampis@redisni.org

Vinh Thong Ta
*Department of Computer Science, Edge Hill University,*
Ormskirk, UK
tav@edgehill.ac.uk

Max Hashem Eiza
*School of Computer Science and Mathematics, Liverpool John Moores University (LJMU)*
Liverpool, UK
m.hashemeiza@ljmu.ac.uk

***Abstract*—With the purpose of defending against lateral movement in today's borderless networks, Zero Trust Architecture (ZTA) adoption is gaining momentum. With a full scale ZTA implementation, it is unlikely that adversaries will be able to spread through the network starting from a compromised endpoint. However, the already authenticated and authorised session of a compromised endpoint can be leveraged to perform limited, though malicious, activities ultimately rendering the endpoints the Achilles heel of ZTA. To effectively detect such attacks, distributed collaborative intrusion detection systems with an attack scenario-based approach have been developed. Nonetheless, Advanced Persistent Threats (APTs) have demonstrated their ability to bypass this approach with a high success ratio. As a result, adversaries can pass undetected or potentially alter the detection logging mechanisms to achieve a stealthy presence. Recently, blockchain technology has demonstrated solid use cases in the cyber security domain. In this paper, motivated by the convergence of ZTA and blockchain-based intrusion detection and prevention, we examine how ZTA can be augmented onto endpoints. Namely, we perform a state-of-the-art review of ZTA models, real-world architectures with a focus on endpoints, and blockchain-based intrusion detection systems. We discuss the potential of blockchain's immutability fortifying the detection process and identify open challenges as well as potential solutions and future directions.**

## I. INTRODUCTION

With the revolution of cloud computing, most businesses' resources and data are no longer stored on premises. Moreover, the recent COVID-19 pandemic has significantly changed work patterns, as most employees and businesses had to switch to working from home. Homeworking (and remote working) open organisations up to new and severe security risks, as many "untrained" employees connect to their work Information Technology (IT) systems with their own devices. Cloud computing and remote working are examples of why businesses must expand their digital security perimeter and adapt to the contemporary trends.

In a traditional perimeter-based security model, the organisation's resources, and assets, inside the perimeter, are assumed to be benign and trusted. Perimeters are usually protected by security measures such as firewalls or intrusion detection systems. This model seems to be less effective in the world of cloud computing and remote working, as indicated by several cyber-attacks (e.g., [1-5]) targeting employees working remotely.

Trust is the fundamental principle a traditional perimeter-based security model relies on. The employees' or collaborators' devices and organisation assets (i.e., endpoints) are typically trusted by default regardless of their condition. If attackers can take control over any of these endpoints, the perimeter is compromised and further access to information and data can be potentially achieved via lateral movement.

Firewalls, antivirus technologies, Intrusion Detection and Prevention Systems (IDS/IPS), and Web Application Firewalls (WAFs), in other words, the big stone walls and armoured front doors, are no longer enough to keep modern IT and Operational Technology (OT) environments safe [6]. Perimeter-based security was the main concept adopted by multiple companies, especially when their data resided in on-premises data centres. The traditional defensive model founded on internal and external disparity is becoming obsolete [7], while at the same time the threat landscape is dramatically evolving [8], ultimately leading to the fall of perimeter-based security architecture.

To cope with today's complex network infrastructures and the current and advancing threat landscape, a new security architecture is needed. ZTA has emerged by establishing a borderless digital identity-based perimeter, where data is at the epicentre of the security architecture and the breach mindset dominates the threat model leading the access control landscape, operations, hosting environments, endpoints, and inter-connecting infrastructures. ZTA fosters a new security architecture in which, by default, any device, system, user, or application should not be inherently trusted based on its location in a network. On the contrary, trust shall always be earned and verified regardless of the location. Nevertheless, this does not necessary mean that in the ZTA context trust is eliminated but should be minimised until proven otherwise via the ZTA tenets and core components.

With traditional perimeter-based defences, determined attackers can still bypass ZTA security health checks if they can establish an authenticated and authorised foothold on the endpoint. For instance, a potential malware in the operating system kernel can tamper with the security checks conducted in the context of a ZTA. This eventually results in bypassing fundamental controls implemented in a ZTA, which would allow attackers to perform several user and device centric malicious activities besides lateral movement. Therefore, an effective intrusion detection approach is required to address the endpoints' vulnerability, which can be seen as the Achilles heel of ZTAs.

In this paper, we aim to examine how ZTA can be augmented onto endpoints using the potential of blockchain's immutability fortifying the intrusion detection process to eliminate the problem mentioned above. To achieve that, we first review the core tenets, capabilities, and requirements of zero trust. Secondly, we categorise existing real-world zero trust implementations and discuss their strengths and weaknesses. Thirdly, we explore the potential of blockchain in developing and improving Distributed Collaborative Intrusion Detection Systems (DCIDSs) that can alleviate the Achilles heel of ZTA (i.e., endpoints' vulnerability). Finally, we discuss the open questions and challenges, as well as highlight potential solutions and research directions to ZTA and distributed blockchain-based IDS.

To the best of our knowledge, this is the first work to shed light on utilising blockchain technologies to successfully augment ZTA onto endpoints. The main abbreviations used throughout this paper are given in Table 1 with their definitions.

*Table 1 – Abbreviations/Definitions.*

| **Abbreviation** | **Definition** |
|---|---|
| APT | Advanced Persistent Threat |
| APTs | Advanced Persistent Threats |
| BNs | Bayesian Networks |
| BYOD | Bring Your Own Device |
| CAML | Correlated Attack Modelling Language |
| CBAC | Context-Based Access Controls |
| CBSigIDS | Collaborative Blockchained Signature-Based Intrusion Detection System |
| CIDNs | Collaborative Intrusion Detection Networks |
| CIDSs | Collaborative Intrusion Detection Systems |
| CIoTA | Collaborative IoT Anomaly Detection |
| CSA | Cloud Security Alliance |
| DCIDS | Distributed Collaborative Intrusion Detection System |
| DCIDSs | Distributed Collaborative Intrusion Detection Systems |
| DDoS | Distributed Denial of Service |
| DISA | Defence Information Systems Agency |
| DLT | Distributed Ledger Technology |
| GDs | Global Detectors |
| IAM | Identity and Access Management |
| IAP | Identity Aware Proxy |
| IDS | Intrusion Detection System |
| IDS/IPS | Intrusion Detection and Prevention Systems |
| IDSs | Intrusion Detection Systems |
| INDRA | Intrusion Detection and Rapid Action |
| IoT | Internet of Things |
| ISS | Information Sharing System |
| IT | Information Technology |
| LAMBDA | Language to Model a Database for Detection of Attacks |
| LB | Local Broker |
| LDs | Local Detectors |
| MCAP | Micro Core and Perimeter |
| MCAPs | Micro Core and Perimeters |
| NGFW | Next-Generation Firewall |
| NGFWs | Next Generation Firewalls |
| NIC | Network Interface Card |
| NIST | National Institute of Standards and Technology |
| OT | Operational Technology |
| P2P | Peer-to-Peer |
| PBFT | Practical Byzantine Fault Tolerance |
| PEP | Policy Enforcement Point |
| PoB | Proof of Burn |
| PoC | Proof of Capacity |
| PoS | Proof of Stake |
| PoW | Proof of Work |
| RADIUS | Remote Authentication Dial-In User Se |
| RBAC | Role-Based Access Controls |
| SDP | Software-Defined Perimeter |
| SIEM | Security Information and Event Management |
| SQL | Structured Query Language |
| SSO | Single Sign On |
| TCP | Transmission Control Protocol |
| VDI | Virtual Desktop Infrastructure |
| VISA | Visa International Service Association Inc |
| VLANs | Virtual Local Area Networks |
| VM | Virtual Machine |
| VPNs | Virtual Private Networks |
| VPN | Virtual Private Network |
| WAFs | Web Application Firewalls |
| ZTA | Zero Trust Architecture |
| ZTAs | Zero Trust Architectures |
| ZTX | Zero Trust Extended |

## II. ZERO TRUST (ZT)

In this section, we provide a brief history of "zero trust" and ZTA, and we discuss the core tenets, core capabilities, models, and existing approaches of zero trust including real-world implementations.

### A. *History of Zero Trust Architecture*

The Jericho Forum in 2004 introduced the idea (radical at that time) of de-perimeterization [3], which subsequently developed into the broader concept of zero trust. The term "zero trust" was coined by J. Kindervag [15] back in 2010; however, the zero-trust concept was present in the cyber security domain before that. The United States Department of Defence and Defence Information Systems Agency (DISA) proposed a secure strategy, named "black core", which was published in 2007 [16]. Black core discussed the transition from a perimeter-based security architecture to one that emphasises on securing individual transactions.

The wide-spread adoption of cloud and mobile computing greatly contributed to the evolving of ZTAs, and as part of it, for instance, approaches such as identity-based architectures slowly gained attention and broader acceptance. Google published a series of six documents under the name "BeyondCorp" on how to achieve a zero-trust architecture [17-22]. The BeyondCorp project advocates for the concept of de-perimeterization, arguing that perimeter-based security controls no longer suffice, and that security should be expanded to users and devices. As a result of this project, Google abandoned the traditional way of remote working based on Virtual Private Networks (VPNs) and managed to provide a reasonable assurance that all corporate users could access Google's network via insecure and unmanaged networks.

### B. *From Traditional Perimeter-based Architecture to ZTA*

As a philosophy, "zero trust" assumes that trust in users, devices, workloads, and network traffic should not be implicitly granted [15] with the consequence that all entities must be explicitly verified, authenticated, authorised, and constantly monitored. One of the core objectives of zero trust is to severely inhibit the ability of adversaries to move laterally, once they successfully manage to compromise a user's device, or even simply steal their credentials. As such, the IT infrastructure needs to be shaped and prepared accordingly.

The traditional perimeter-based security architecture creates multiple zones of trust [2]. Not all zones adhere to the same rules or to the same level of trust. In fact, users might not be able to even reach into the next zone if not explicitly allowed by the relevant component. This is referred to as defence-in-depth, as discussed by Smith [23] and depicted in Figure 1 or as a castle-and-moat approach [24]. Note the different zones (Internet, demilitarized zone, trusted, and privileged) are being protected by various perimeter-based controls such as a local broker, a VPN gateway, multiple firewalls, and application services prior to reaching the mainframe. In this example (i.e., Figure 1), the mainframe is a core banking system, responsible for all transactions hence it is separated entirely in a privileged zone.

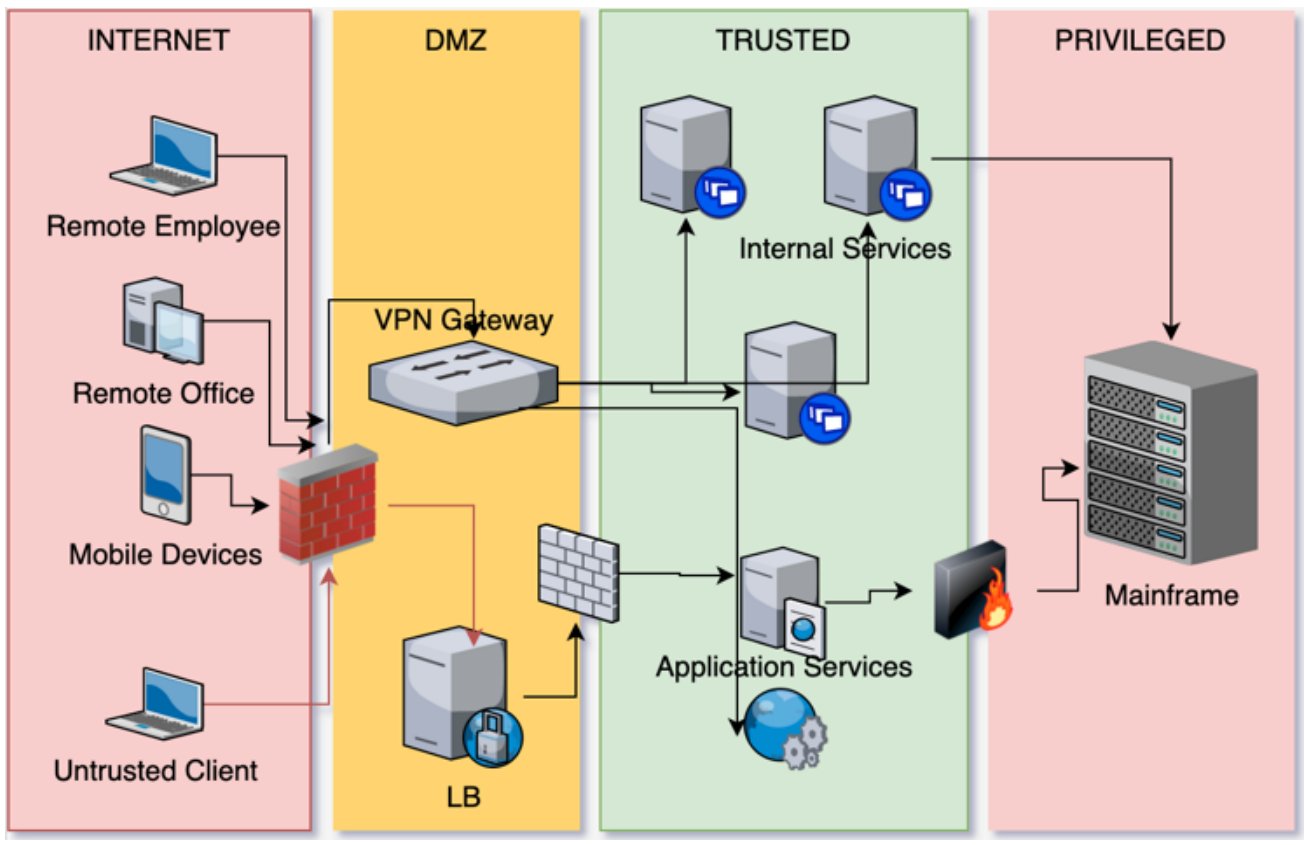

*Figure 1 – A traditional security architecture.*

Unlike a traditional security architecture, zero trust calls for thinking, building, and protecting from the inside out. Based on the above-mentioned works from Google [19] [20], Jericho [3] and Kindervag [15] [25], there is one immediate and important observation that in ZTA the VPN technology can be eliminated once the network locality dependency becomes irrelevant. VPN, in short, allows a user (denoted by "Remote Employee" in Figure 1) working remotely, to connect to an office (denoted as "TRUSTED" in Figure 1), via a secure encrypted channel. However, the endpoints should be protected by other means since VPN encryption only addresses the tunnel between the "Remote Employee" and the "TRUSTED" zone. When the "Remote Employee" is authenticated and the tunnel is successfully established, he/she receives an IP address in the remote network of the "TRUSTED" zone. On that tunnel, the traffic from the "Remote Employee" to the "TRUSTED" zone is decapsulated and routed, therefore, leading to an "official" backdoor. Moreover, the single-entry point denoted as "VPN Gateway" acts as a single point of failure or strangle point for the architecture and the network. Hence, if we start considering the network location as irrelevant, while at the same time applying a proper set of controls, then VPN can be eliminated if there are no further dependencies (e.g., apps with legacy protocols). That said, authentication and authorisation alongside policy enforcement should immediately move closer to the network edge and endpoints.

To reflect the arguments above, we draw Figure 2 that shows a reference to ZTA. For the sake of simplification, in the figure, we include only the core components, for instance, a Local Broker (LB), the remote employees, mobile devices, untrusted clients, and numerous services that require protection. Compared to the perimeter-based architecture shown in Figure 1, there are no zones, and the security is being built from the inside out. In addition, there are neither VPN gateways, nor firewalls to filter network traffic, and most importantly there is no single gateway of entrance. We notice; however, a policy enforcement point at the control plane. This ZTA reference does not create any strangle point like in the case of the perimeter-based architecture.

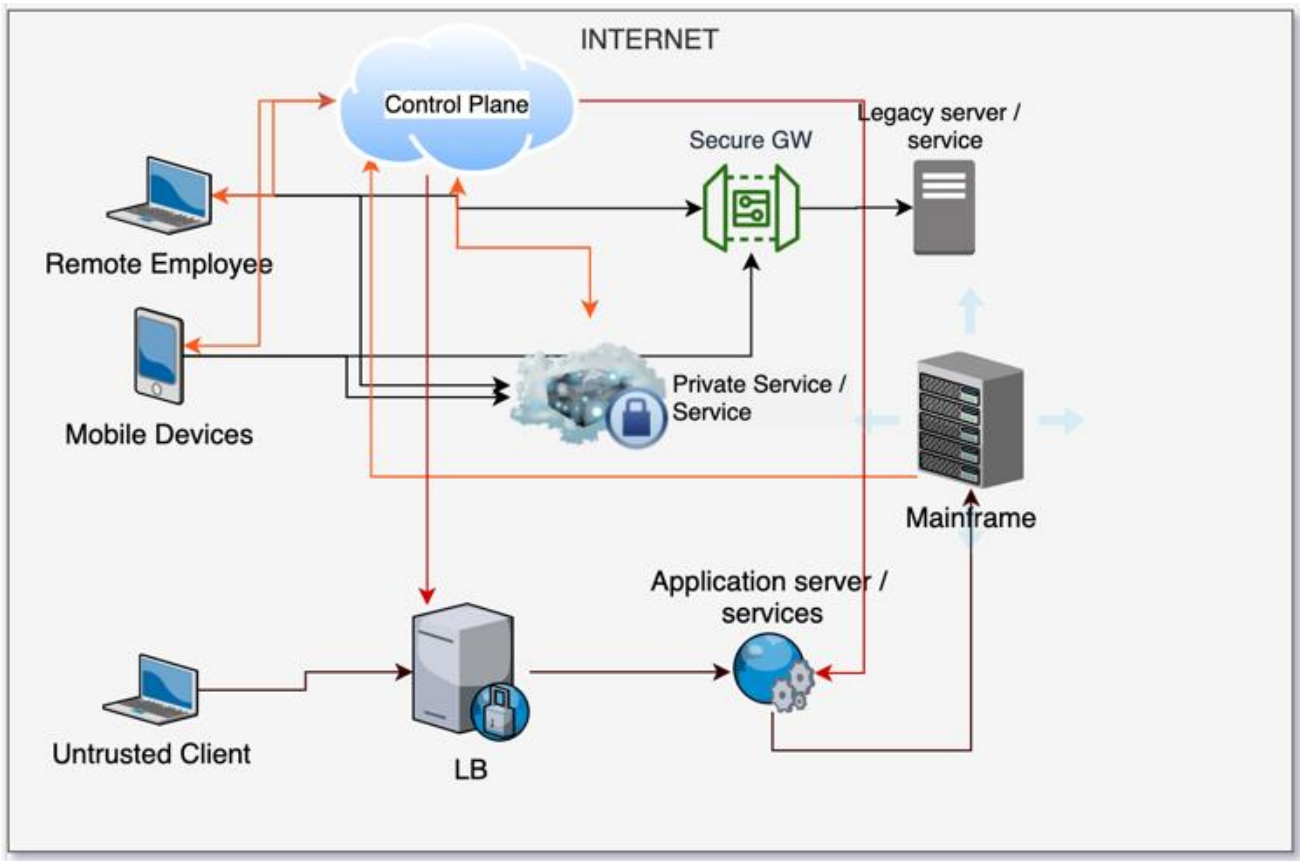

*Figure 2 – A high level ZTA reference.*

To make this ZTA reference vendor agnostic, we simply use the generalised term of control plane, and distinguish between control plane and data plane. This is a known concept in cloud architectures, and we use the same analogy here to leverage the fact that the control plane poses inherent and unlimited access to the data plane. All access requests to resources must be directed through the control plane, where a set of authorisation and authentication policies, rules and context parameters must be met. Access to more private resources (e.g., a payment router or a mainframe resource) can be further restricted based on Role-Based Access Controls (RBAC) enhanced by Context-Based Access Controls (CBAC) on the same level. Finally, if the control plane concludes that the request should proceed, then it coordinates and configures as necessary the data plane to accept the connection from the requestor. Additionally, the control plane can potentially coordinate the setup of an encrypted tunnel for the requestor and the destination resource.

### *C. Zero Trust Core Tenets*

In this section, we review the five core tenets of zero trust, based on the works of Rose et al. [7], Gilman et al. [2] and Jericho [3], as follows:

#### *1) Access Segmentation*

Every access to a resource must be appropriately segmented, in order that no single entity can access the entire network or even a large part of it. Furthermore, a minimum number of entities must be able to explicitly access critical data. This explicit access applies particularly to administrators, where in most cases they tend to preserve unlimited and uncontrolled access throughout the whole network.

#### *2) Universal Authentication*

All entities, including users, devices, applications, and workloads, having any form of interaction with the corporate network must be authenticated regardless of their location in the network.

### 3) *Encrypt as Much as Possible*

ZTA assumes a breach (i.e., the worst-case scenario), therefore, the network is always considered hostile, and trust cannot be inherently granted. That said, one must always assume that a potential adversary can intercept any type of communication happening throughout the network. As a result, all communications should be end-to-end encrypted externally or internally.

### 4) *The Principle of Least Privilege*

All entities in a ZTA must be restricted to the least amount of privilege required for that specific entity to complete its mission or operation. This includes, for instance, what an entity can access, and where and for how long. Moreover, the overall trustworthiness of an entity must be evaluated based on the context or attributes, ultimately indicating if it shall be trusted or not.

### 5) *Continuous Monitoring and Adjusting*

Every entity (internal or external) in a ZTA should be monitored. In this context, all network traffic, system events, and access attempts should be monitored and recorded regardless of failure or success. These must be continuously analysed and cross-checked against the security policy. The outcome should be then used to adjust the relevant policies when needed.

Jointly, these five core tenets form the concept of zero trust. Although the above-mentioned papers can be found with slightly different titles or descriptions, they share the same essence. Those principles must be applied at many distinct levels, for instance, users as well as administrators, and on many different domains, such as traditional networks as well as on cloud infrastructures. It needs to be highlighted that, although zero trust is gaining momentum and the market for the related products are expected to double by 2024 [28], there is limited vendor agnostic, scientific critical literature available.

## D. *Zero Trust Core Capabilities*

In this section, the core capabilities of a ZTA are presented based on the National Institute of Standards and Technology (NIST) special publication 800-207 [7], Google's BeyondCorp [20] and Kindervag et al. [15]. The core capabilities include network and system access control, traffic filtering, application segmentation and execution control, operational analysis, and policy enforcement.

### 1) *Network Access Control*

Network access control says that the authentication of all entities should happen before allowing entities further access to organisational assets. This can be achieved by proper network segmentation and a robust access control policy.

### 2) *System Access Control*

This category of capabilities deals with the file and user access controls. These can be implemented by using login agents and different cryptographic controls, such as full disk encryption.

### 3) *Traffic Filtering*

This category of capabilities is about the enforcement of network segmentation and prevention of unauthorised connections. For this purpose, firewall technologies along with IDS/IPS and traffic analysis tools can be applied. In addition, monitoring of unusual traffic behaviour should be implemented.

### 4) *Application Segmentation*

Like network segmentation, applications must be isolated from each other, and user access should be explicitly limited to only those applications users need to successfully perform their duty.

### 5) *Application Execution Control*

This deals with the prevention of unwanted, potentially malicious, applications that have not been previously authorised and approved to be executed. Application whitelisting is a common control for this category.

### 6) *Operational and Forensic Analysis*

This deals with analysing the systems and resources for evidence of breach or to detect anomalies. The most common technical approaches that support this include (i) host-based intrusion detection systems, (ii) application monitoring, (iii) forensic tools, (iv) honeypots/honeynets, (v) vulnerability scanners, (vi) penetration testing, (vii) threat intelligence, and (viii) red teaming. In addition, Security Information and Event Management (SIEM) tools, as well as Advanced Persistent Threat (APT) detection and prevention methods have been widely used to tackle more advanced threats.

### 7) *Policy Engine / Policy Enforcement*

This includes vulnerability analysis and prioritisation, operational risk, and behavioural analysis. To help readers understand the connection among the core capabilities, in Figure 3, we draw a typical application of the seven capabilities in an example notional bank's information technology architecture. In this figure, the green stickers highlight the measures to satisfy the zero trust core capabilities and core tenets.

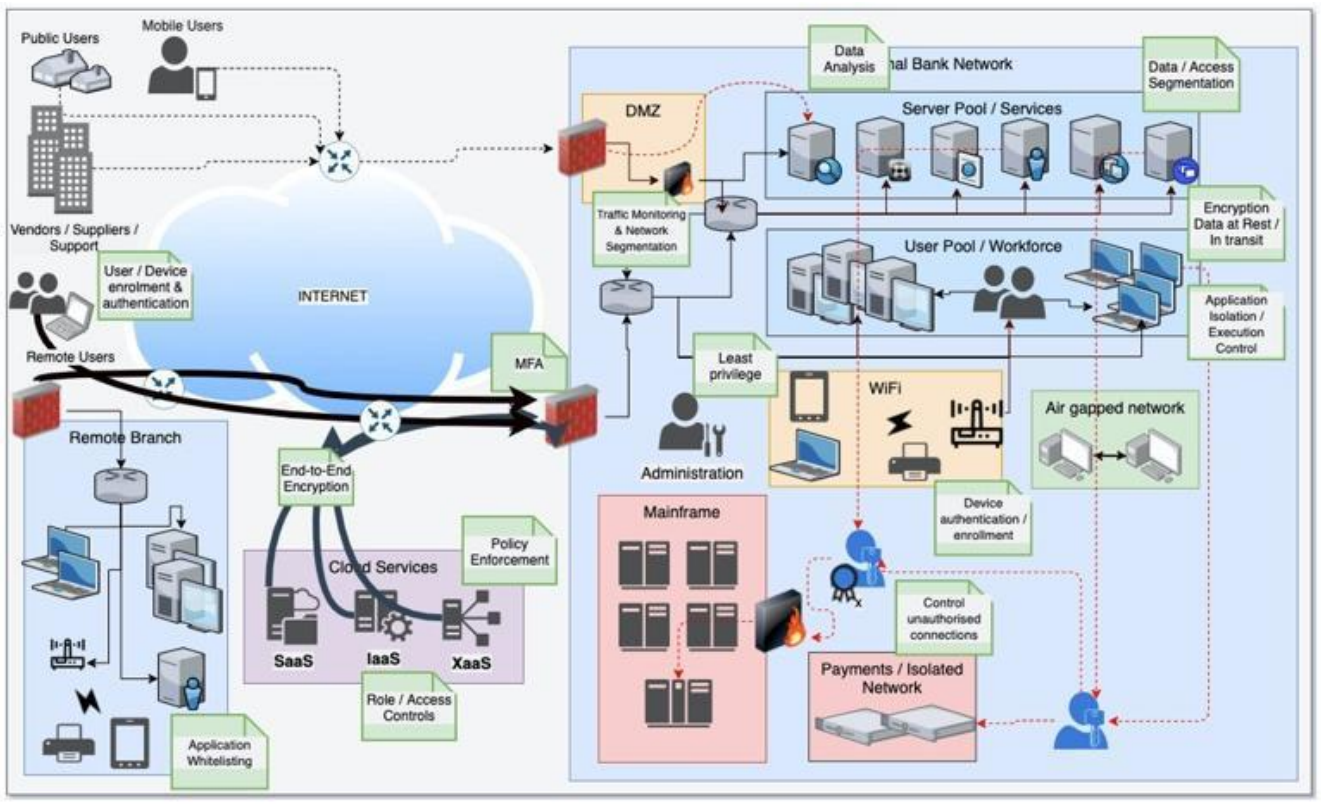

*Figure 3 – An example ZTA capabilities reference.*

## E. *NIST Zero Trust Models*

In the following, we discuss three zero trust deployment models, presented in the NIST standardisation document [7]. These deployment models are high-level concepts, without any real-world implementation examples. Each model is composed of a control plane and a data plane. The control plane includes the policy engine and policy administrator, while the data plane contains the components that support data transmission. Note that the core tenets and capabilities outlined in the previous two subsections can be implemented as part of each high-level deployment model.

*1) Device Agent / Gateway-Based Deployment*

In this deployment model [7], as shown in Figure 4, the Policy Enforcement Point (PEP) must be highly integrated with two major components, the endpoints, tagged as 'Enterprise System' (which can be laptops, PCs in a remote location, or handheld devices), and the resource or application(s) that is subject to a user access request.

To implement this model, an agent is required to be installed on the endpoints. This model provides the best overall control among the three models, because the agent acquires real time contextual information of the resources the users are trying to access for the endpoints and the users, at any time. As a result, a decision by the control plane can be made at any point and the necessary configuration of the data plane is instant and highly accurate.

Nonetheless, a drawback of this model is the overhead that comes with the agent installations and the full integration of the data resource with the gateway. A good example of this model is the Google's BeyondCorp implementation [5].

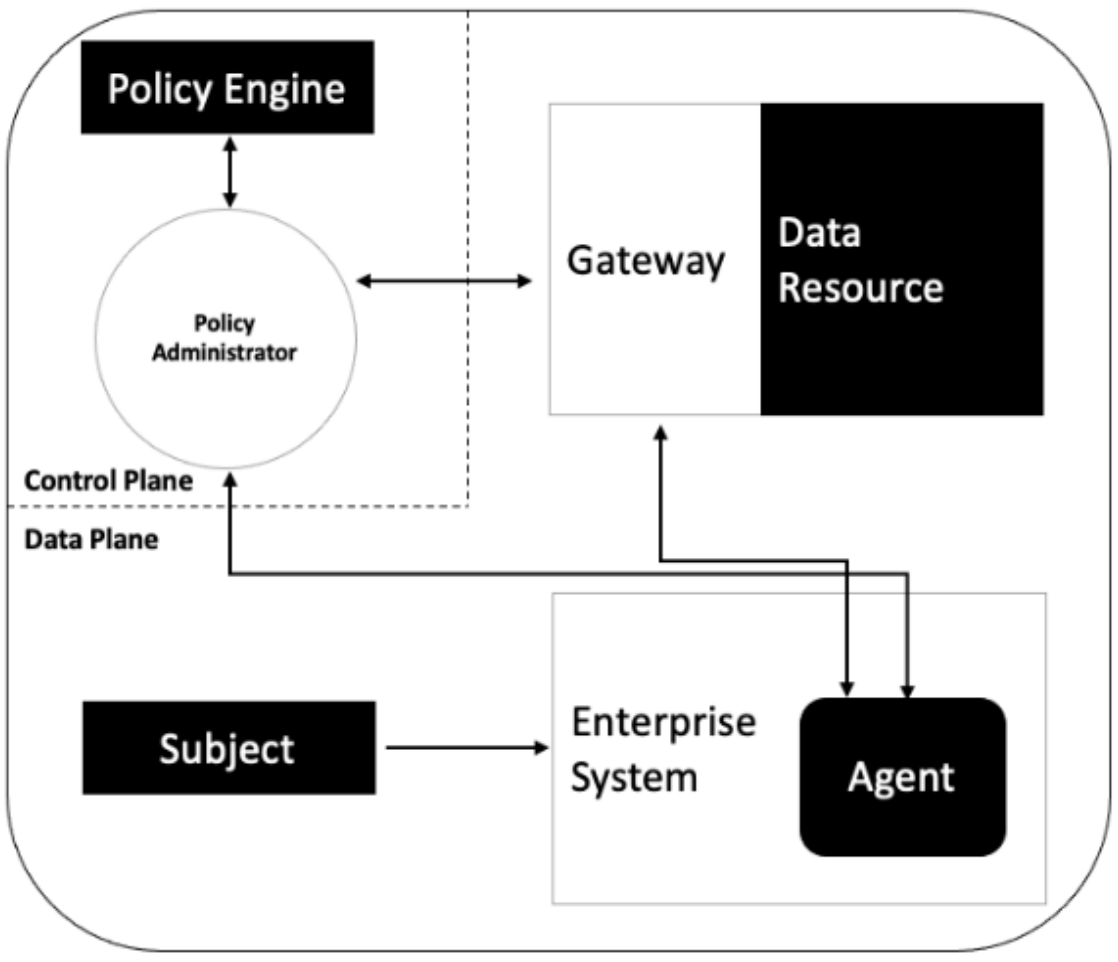

*Figure 4 – NIST Device Agent/Gateway-Based Deployment [7].*

*2) Enclave-Based Deployment*

Like the previous case, this model [7] again requires an agent to be installed on the user's endpoint, however, the PEP is placed in front of an enclave of resources. Unlike the first deployment model, there is no requirement for a tight integration between the resources, which is one of the advantages of this model as shown in Figure 5. A disadvantage, however, is that a zone of implicit trust is automatically created amongst the gateway and the resources, and therefore, the advantage that comes with the acquired contextual information, as seen in the first model, is lost.

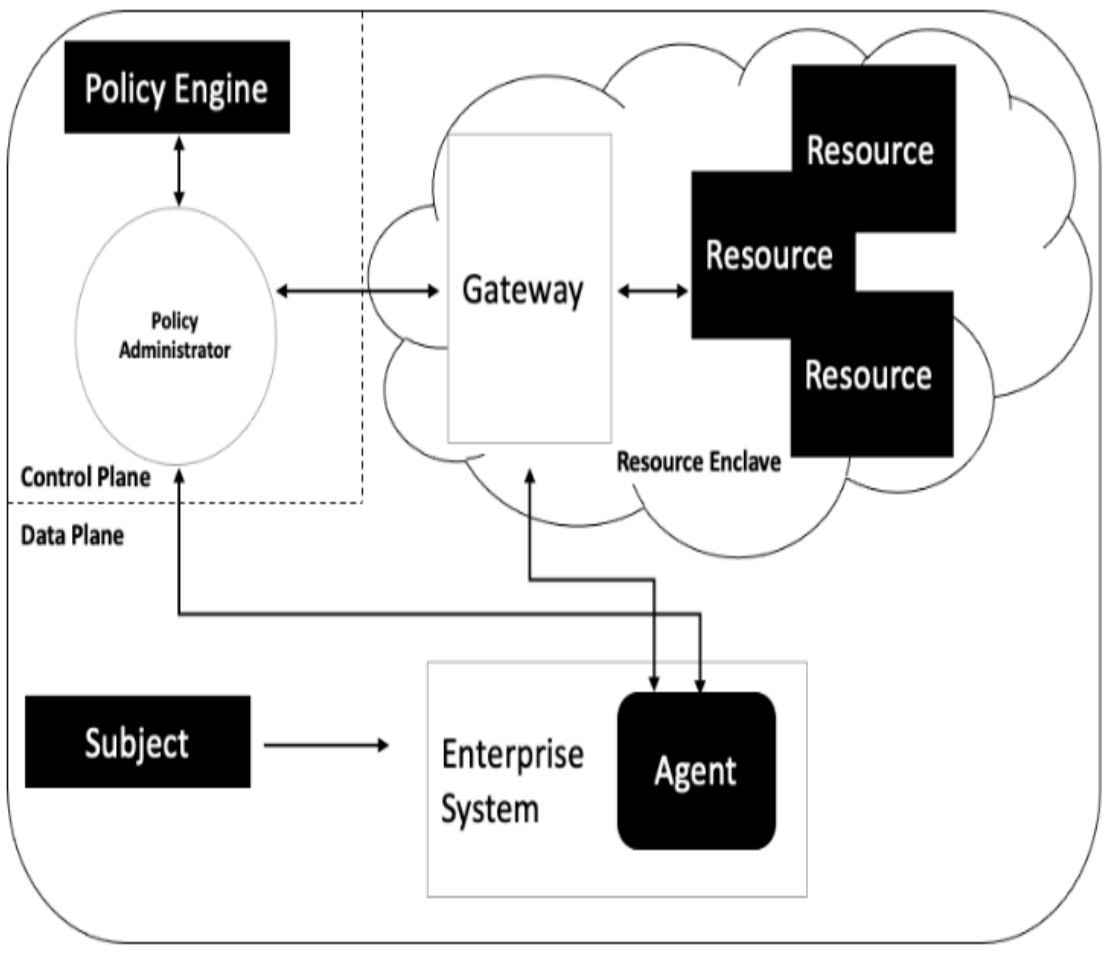

*Figure 5 – NIST Enclave-Based Deployment [7].*

*3) Resource Portal-Based Deployment*

In this model, the PEP is neither integrated with the user endpoint nor the application or service [7], as shown in Figure 6. A gateway is positioned accordingly in the network corridor, and responsible for controlling access to the subject resources. The advantage of this deployment model is that it is agentless, namely, no special software is required to be installed on the user's endpoint(s), and the subject application(s) / resource(s) do not require any modifications. However, its drawback is the loss of fine-grained access control towards the resources or applications, and hence, limiting zero contextual information that can be used to make context aware decisions. The first example of this model was presented by Forrester [15] utilising technologies such as Virtual Local Area Networks (VLANs) and Next Generation Firewalls (NGFWs) to achieve segmentation (we will discuss more about it in section II.F).

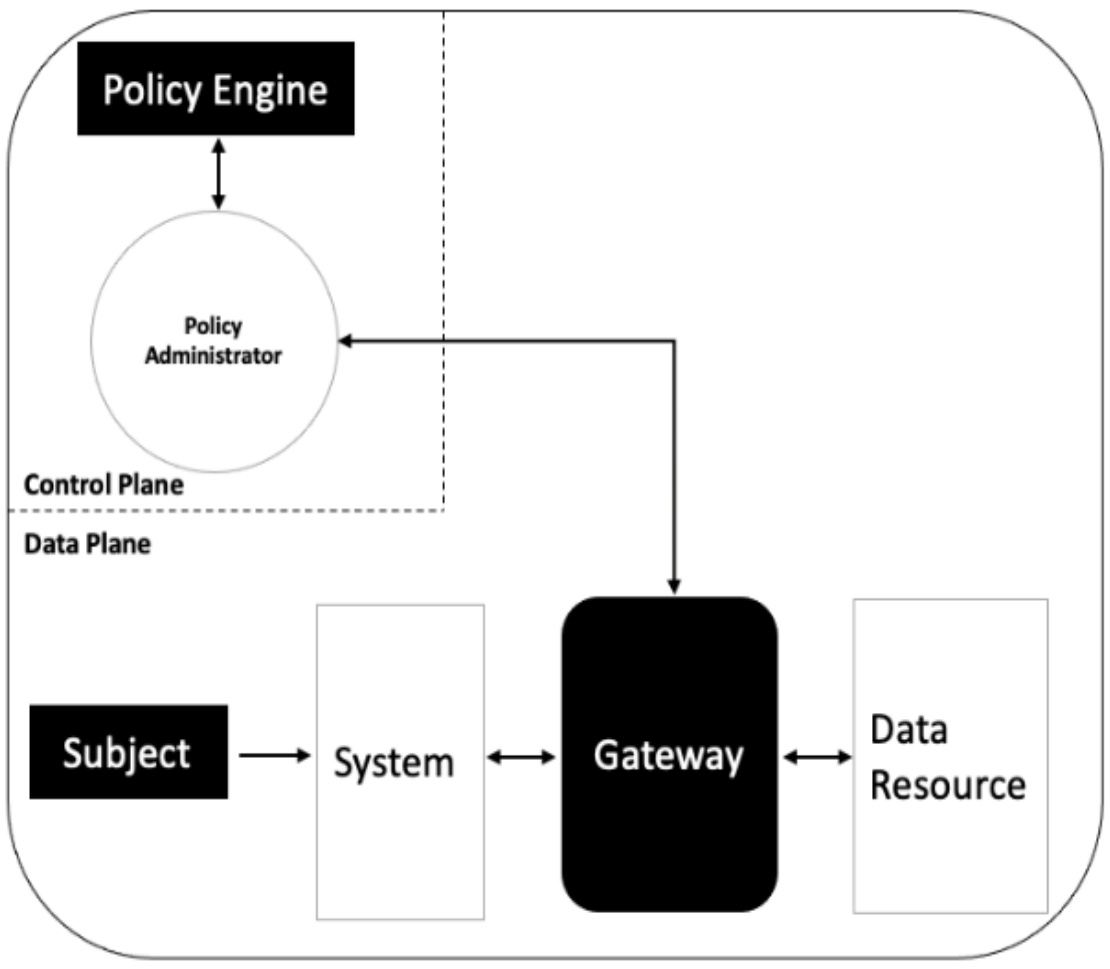

*Figure 6 – NIST Resource Portal-Based Deployment [7].*

To conclude this section, in Table 2, we provide a comparison of the three zero trust deployment models based on the four discussed characteristics, alongside their advantages and limitations.

*Table 2 – Advantages / Disadvantages & Attribution Table of NIST's ZT deployment models.*

| NIST Deployment Model | PEP Location | Agent Required | Control/ Data plane Integration | Contextual information / fine grained access controls | Advantages | Limitations |
|---|---|---|---|---|---|---|
| **Device Agent/Gateway -Based** | Attached to resources | System & resource | Tight | Highly available – yes | A context aware environment can be introduced | De facto requirement of agent installation |
| **Enclave-Based** | In front of resources | System | Medium | Limited availability – not possible | There is no need for tight integration between resources | The introduction of a context aware environment is lost |
| **Resource Portal-Based** | In between system & resources | None | Loose | Limited to zero – not possible | It is agentless | Loss of fine-grained access controls towards the resources or applications |

### *F. Zero Trust Architecture Approaches and Implementations*

In this section, we discuss the existing approaches and implementations for ZTAs. First, we discuss the more theoretical approaches and concepts proposed in research papers. Afterwards, we present some important real-world ZTA implementations by enterprise. At the end of this section, we summarise and compare the real-world implementations based on the NIST deployment models in Table 3.

#### *1) Theoretical Approaches for ZTAs*

Cloud and mobile computing introduced and enabled borderless networks; therefore, it is imperative to re-design cyber security controls accordingly and not just focus on the corporate perimeter. DeCusatis et al. [26] identified the limitations of the existing best practices regarding network segmentation. Grounded on a steganographic overlay, they discussed a novel architecture as an enabler to a zero-trust approach. Technically, the so-called steganographic overlay embeds authentication tokens within the first-packet authentication and Transmission Control Protocol (TCP) requests. An experiment deployment was demonstrated in both the traditional and cloud computing environments.

Rose et al. [7] first provided an abstract definition of ZTA, while also contributing to the common body of knowledge by specifying general deployment models and use cases where ZTA could enhance an overall cyber security posture of an enterprise. Embrey [29] identified the top three factors driving the adoption of ZTA and stressed its necessity to enhance security and policy controls at both the user's and device's level. Mehraj and Banday [30] proposed a conceptual zero trust strategy, explicitly designed for cloud environments. Their efforts also emphasise trust establishment and the further trust challenges applicable to cloud computing. Yan and Wang [31] performed a survey on zero trust components and the key technologies for ZTA. They also applied some of the subject technologies and related them to specific scenarios, to highlight further the advantages of ZTAs.

Keeriyattil studied the whitelisting approach [32], at the network level. The ingress and egress traffic of a virtual Network Interface Card (NIC) were examined against a given list of firewall policies. Based on the whitelisting concept, if no matching rule is found for a specific traffic flow, then the packet is simply dropped. Using specific technologies (e.g., VMWare NSX) the author demonstrated how only the traffic that is checked against specific records would be allowed. Mital [33] discussed the features of the Distributed Ledger Technology (DLT) and blockchain technology that would be applicable to the zero-trust context. Specifically, the author discussed how the immutability property of blockchain could help in establishing higher integrity standards. In addition, the elimination of a possible single point of failure in ZTA could help with maximising the availability of the system/network, due to the "inherent" relevant attributes of DLT.

#### *2) Real-world ZTA Implementations*

In this section, we review four relevant "real-life" ZTA approaches, namely, Google's BeyondCorp [17], Forrester NGFW/ZTX [25], Cloud Security Alliance (CSA), Software-Defined Perimeter (SDP) [34], and VMWare NSX [32]. Those architectures are the current dominating real-world deployment models [35], unlike the previous high-level architectures from NIST in section II.E.

##### *a) Google's BeyondCorp*

Following a hacking campaign by the Anonymous group named Operation Aurora in 2009 [36], Google launched the BeyondCorp project. A detailed report published by McAfee labs on the lessons learned from Operation Aurora [37], noted that the attackers were able to access the internal network by specifically targeting the sources of intellectual properties and using the compromised system as a starting point (also known as "jump-point") to move laterally.

Consequently, Google's primary goal was to remove the inherent trust acquired by its users and devices, due to their placement (physical or electronic) within the corporate network. Moreover, in case a user or a device was compromised, as seen during Operation Aurora, a secondary goal was to minimise the probability of an adversary moving laterally through the network and compromising further entities. Three core tenets were the derivative of the first whitepaper of BeyondCorp in 2014 [17]:

- The services that a user/device can access must not be determined by a specific connection and especially the location of the connection.
- All access to services must be determined based on contextual information.
- All access to services must be authenticated, authorised, and encrypted.

Figure 7 highlights the access and traffic flow alongside the components of the BeyondCorp zero trust implementation. The components include the access proxy, the access control engine, the pipeline that receives input from the device inventory database, the user/group database, and finally, the trust inference alongside the certificate issuer. Such an approach can be mapped back to the Device Agent/Gateway-based deployment model proposed by NIST (see section II.E).

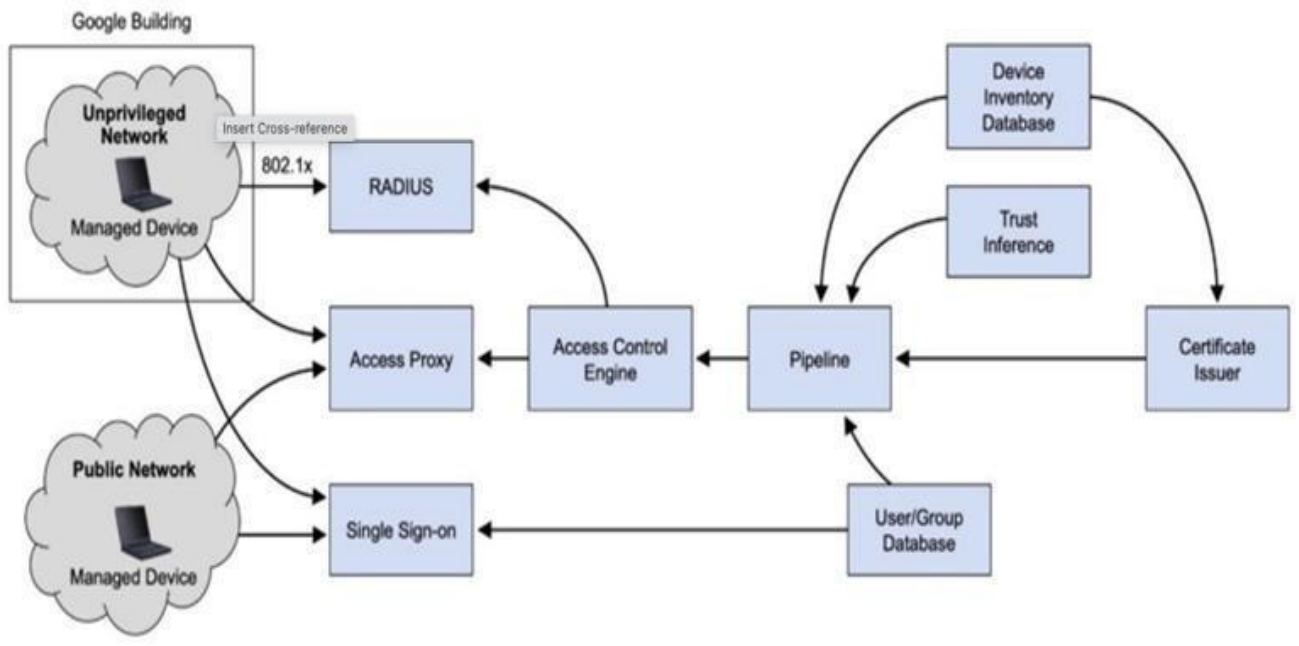

*Figure 7 - BeyondCorp Traffic/Access Flow & Components [17].*

Note that in this model, the public and the internal networks inside a Google's building have absolutely no differences when it comes to user and device privileges as both are considered unprivileged. Device authentication on the internal unprivileged network is performed via the 802.1x standard through a Remote Authentication Dial-In User Service (RADIUS) server. Prior to accessing that network, all users follow the same flow through a Single Sign On (SSO) mechanism, which provides authentication to resources. Complementing this zero-trust model, an innovative element is their Identity Aware Proxy (IAP), which works synergistically with context-based access control. The access to resources is not implicitly allowed for the user/device being simply part of the corporate network. Quite the reverse, access is explicitly granted based on context and policy.

The BeyondCorp model authenticates the users on the application layer of the network. There is a heavy reliance on this aspect since most of their applications and services are web-based. Furthermore, as Google applications are mostly developed internally, combined with their own existing SSO system, this has led to a successful implementation of the new architecture. However, companies without heavy internal development or heavy reliance on web-based services, will probably require a different model. Google has since productized BeyondCorp's evaluated model as BeyondProd, which is a cloud native security solution [38].

Overall, if an organisation has multiple publicly exposed services with several cloud-based applications accessed by public users, then this is likely to be a suitable model. However, we note that Google only applies this on their cloud infrastructure and, to the best of our knowledge, currently no other organisation offers a similar solution. As a result, applying the BeyondCorp model for a non-cloud environment is not straightforward, and the relocation of several core management controls may be required.

*b) Forrester (Zero Trust Extended) ZTX*

In this model, as depicted in Figure 8 [15], a centralised segmentation engine manages and isolates the enterprise network into multiple Micro Core and Perimeter (MCAP) segments, when and where appropriate. As such, it can enforce traffic rules in between MCAPs. Figure 8 shows the NGFW being used as a segmentation engine to form multiple MCAPs. Such an approach can be mapped back to the "Resource Portal" model outlined by NIST (see section II.E).

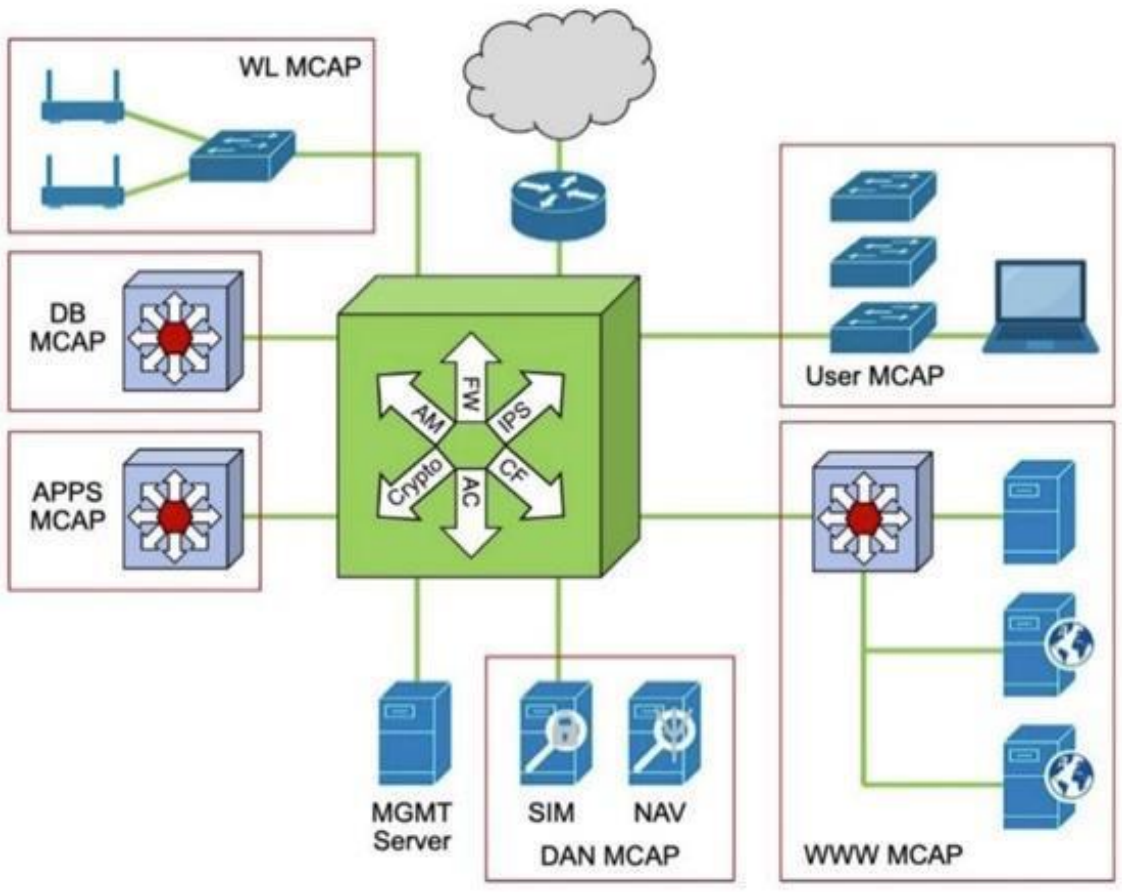

*Figure 8 - Forrester's NGFW used as a segmentation engine forming MCAPs [15].*

As highlighted in Table 2 in reference to the Resource Portal model, the required changes in components for this model prior to implementation are minimum or near zero, hence, it can be an attractive choice. However, this model makes use of the information available in the data packets to enforce trust. This approach is less "granular" compared to the architectures that integrate tightly with endpoints and services. Another drawback of this approach is that users cannot be directly authenticated with the NGFW segmentation engine. More specifically, the segmentation engine is not capable of enforcing policies based on the contextual information of users and devices.

Many organisations are already deploying a resource portal architecture, which can be seen as a good match for this ZTA. This architecture alongside the enclave-based, is likely to be the best for, and the easiest to deploy in, a Bring Your Own Device (BYOD) or an Internet of Things (IoT) environment, because the devices can be placed within their own enclave or MCAP. However, an important shortcoming is that the access control mechanism in this model can be less fine grained than in other architectures. In addition, there is a dependency on further integration with other technologies such as Identity and Access Management (IAM), device management systems or VPNs, to achieve the same security levels as other architectures.

*c) CSA's Software Defined Perimeter*

The concept of SDP was introduced by a non-profit organisation called the CSA in 2013 [34]. Since then, several SDP based solutions have been developed, and have been proven for large organisations holding its fair share in the market. Using the NIST high-level models to conduct a mapping, SDP would match the Enclave-Based Deployment Model. Namely, an agent is required to be installed at the endpoint and the service, however, there is no integration with the target resource or the target application. Therefore, the agent itself can be taking on the role of a gateway on the service side.

We can find some similarities between this model and the Forrester ZTX approach. For instance, like the NGFW solution described in the previous point, the SDP approach performs network segmentation as a central firewall. It undertakes the role of an overlay network beyond the current network infrastructure. User authentication and identity verification happen at the SDP server, therefore, instantly creating a VPN tunnel between the subject resource and the authenticated user. Figure 9 shows the described SDP controller connection handling process. As can be seen, the workflow is split into control and data channels, and eventually results in a direct VPN tunnel between SDP hosts.

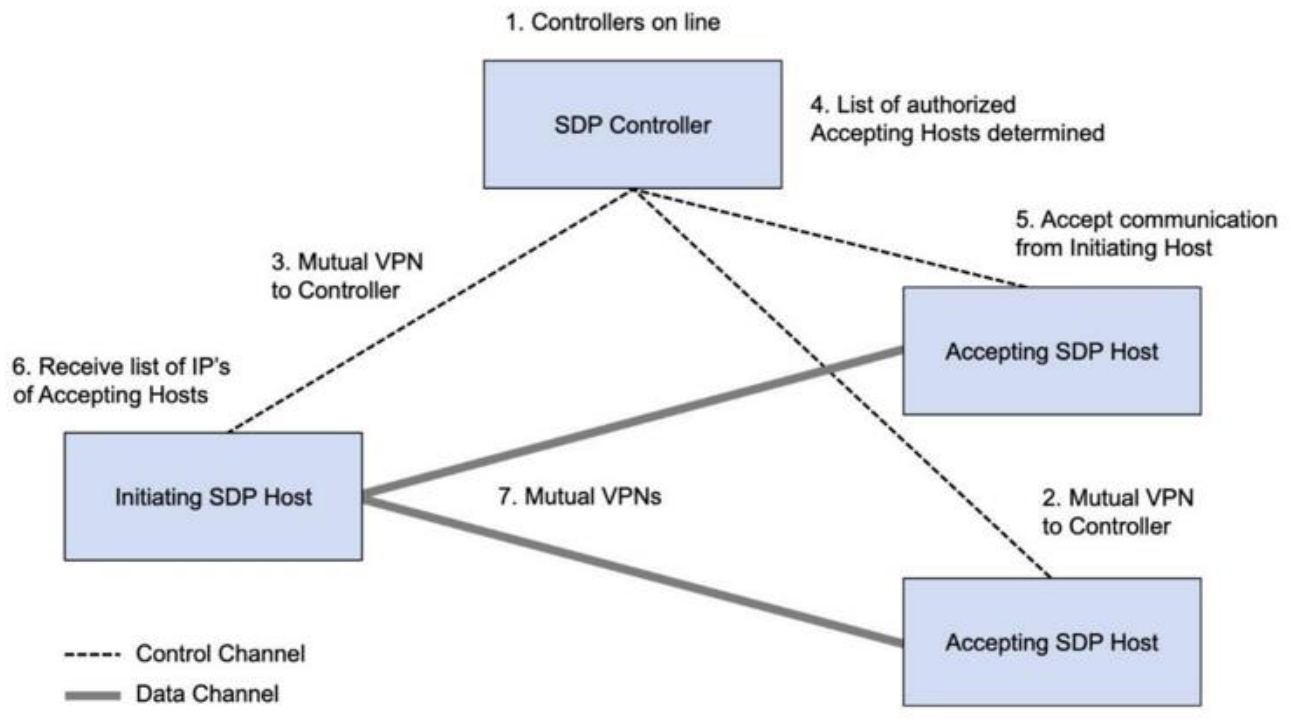

*Figure 9 - SDP Reference Workflow [34].*

The key difference, however, relies on how a VPN and the SDP approach manage and establish the overall trust towards users and devices. For instance, in case of VPN, once a user and/or a device is authenticated and authorised, he/she can access most of the network with trust being implicitly applied by default considering the network location. On the other hand, once a user and/or a device authenticates itself with the SDP controller, a set of role-based access, attributes, and context of user trust is enforced. An important advantage of SDP, nonetheless, is the elimination of the integration with the subject resource (or application). At the same time, installation, and configuration on both the resource and endpoint are still required.

Conclusively, SDP is a new concept being continuously improved, and the relevant market offerings are not yet mature enough, at least at the time of this writing, though they have reached a point where enterprise adoption can be achieved with no significant issues or complications. Moreover, SDP does not require a costly integration with the applications, due to its inherent architecture principle. Finally, SDP can be seen as a perfect match for organisations with multiple IoT systems, or operational technology in general since the gateway can act on behalf of the mentioned devices. Barcelo et al. [39] and Anggorojati et al. [40] confirmed this via the SDP and IoT/OT integration and heavy testing.

*d) VMWare NSX*

The deployment based on VMWare NSX is another real-world ZTA deployment. However, this model is mainly referring to organisations that already leverage the Virtual Desktop Infrastructure (VDI) [32]. The model matches the Device-Agent/Gateway Deployment model, although it assumes that all resources are based on virtualised systems, namely, the applications are hosted on virtual servers. A reference zero trust architecture using NSX is shown in Figure 10.

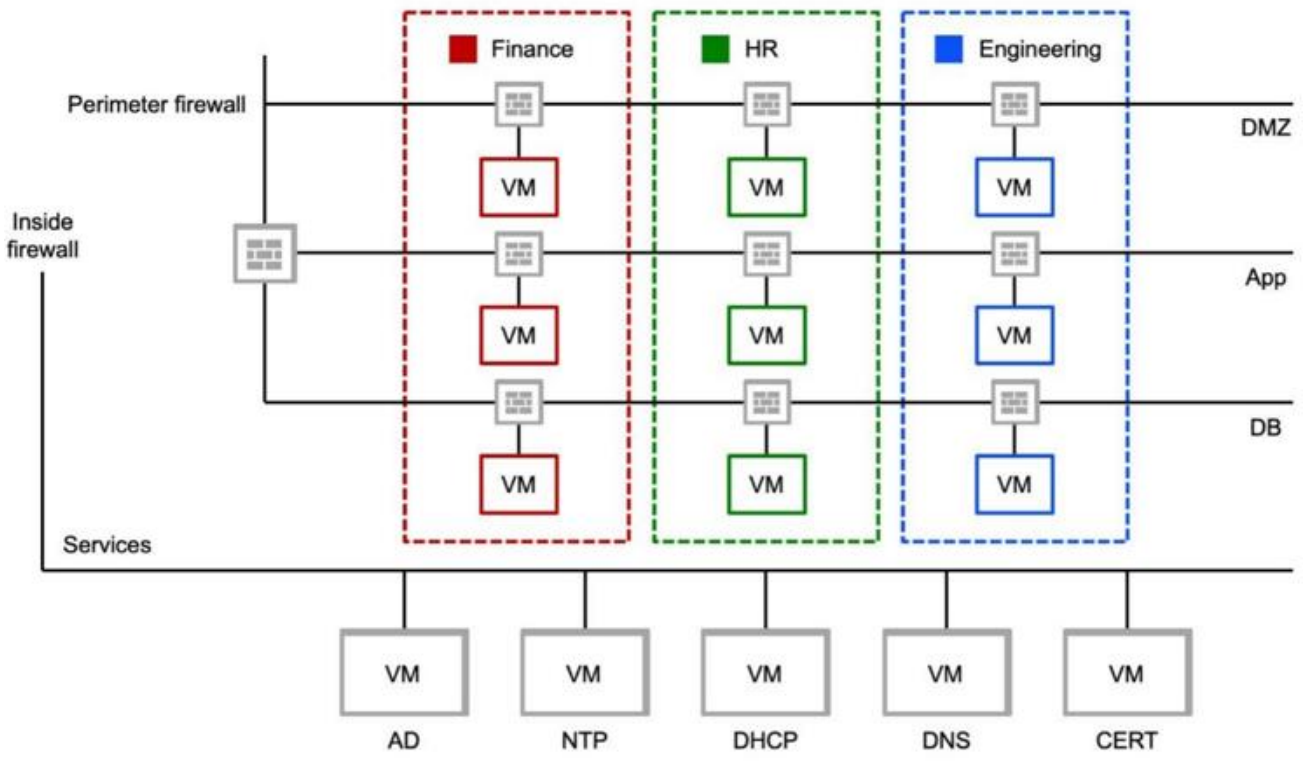

*Figure 10 - Reference ZTA using NSX [32].*

As depicted in Figure 10, the workflow of this architecture starts with a user authentication step on the VDI server. Thereafter, a remote session on a virtual desktop is established and presented to the user. The virtual server and the virtual desktop are the two core components of the NSX based approach. In this case, NSX acts as a firewall where policy decisions and trust management are performed and enforced throughout the network as a whole and in multiple points. Hence, the administrative team can perform access control fine graining in manifold segments, which can be also referred to as micro-segmentation [41].

A major advantage of this approach is the concept of the virtualised desktop. Particularly, the administrator group, who control the full Virtual Machine (VM) or virtual desktop fleet, could refresh or rebuild it on a frequent basis (e.g., at night). Therefore, if we assume an adversary compromising an endpoint via one of the most common adversary methodologies, such as phishing or spear phishing etc., establishing a persistent foothold would be highly unlikely. Hence, this approach would disrupt the so-called cyber kill chain [42] at an exceedingly early stage. On the other hand, most organisations are already deploying a highly virtualised model, but switching into a VDI-based architecture would be costly. In contrast to the SDP approach, this model may be a bad choice for IoT systems due to the virtualisation requirement in the sensors and OT.

*Table 3 - Real-World ZTA implementations mapped to NIST deployment models.*

| **NIST Deploym** | | | | **Contextual information /** | **Real-World** |
|---|---|---|---|---|---|

| ent Model | PEP Location | Agent Require d | Control/ Data plane Integration | fine grained access controls | Impleme ntation |
|---|---|---|---|---|---|
| **Device Agent/G ateway-Based** | Attached to resources | System & resource | Tight | Highly available – yes | **Google's BeyondC orp & VMWar e NSX** |
| **Enclave-Based** | In front of resources | System | Medium | Limited availability – not possible | **Software Defined Perimete r** |
| **Resource Portal-Based** | In between system & resources | None | Loose | Limited to zero – not possible | **NGFW / Forreste r ZTX** |

Finally, building upon Table 2, we map the real-life ZTA implementations to the NIST deployment models, and provide Table 3 above with summarised information.

## III. Open Questions And Challenges to ZTA: The Endpoint Problem

As we mentioned before, the primary goal of ZTA, if properly implemented, is to perform a fine-grained identity-based access control [7] that can specifically prevent the increasingly severe risk of lateral movement. There are multiple access control types such as role-based and attribute-based access controls, however, ZTA performs access control on the identity of the user (i.e., identity-based access control). Moreover, the zero-trust approach primarily focuses on protecting assets, network/user accounts, workflows, and services rather than network segments. The location of the network (e.g., home, work, or a public place) is deemed irrelevant within the ZTA context and its relationship to the overall security posture of the resource.

However, the above argument comes with a fundamental assumption that the core components of a ZTA should be able to contextualise users access requests before granting them access to enterprise resources. Namely, before a user is granted access to corporate resources, several conditions must be met, such as the operating system version, software patch levels, IP address or source/origin, the time of a request (e.g., is it between 09:00-17:00?). Such information is of course subject to each corporate policy and the context. This approach can be effectively implemented if, for instance, we assume extremely locked-down devices, or fully managed devices like in BeyondCorp [17], where only corporate Google Chromebook devices are granted access, without support for the BYOD capability [17].

It should be noted, however, that currently most enterprises run Windows as their core operating system [43], and may run a wide variety of legacy, outdated applications and/or middleware increasing their security risks. Determined attackers have previously demonstrated how the traditional perimeter-based defences can be bypassed, for example, with malware and phishing attacks, to gain a foothold in enterprise networks. Once a device is compromised, the operating system (and the device that runs it) can no longer be trusted, since a potential malware in the operating system kernel can tamper with the ZTA security health checks, which are part of the context built by ZTA. This eventually results in bypassing the fundamental control implemented in a ZTA.

As a result, enterprises that implement one of the current ZTA models might mistakenly trust user devices (or endpoints), as attackers are still able to compromise those devices, and thereafter, ride the already authenticated user's session to perform several user and device centric malicious activities other than lateral movement. A good example is the attack approach like the MITRE ATT&CK navigator [44] that includes malicious payload execution, privilege escalation, and defence evasion to compromise user devices. In case the compromised device belongs to an administrator, the inherent impact of such a scenario is of critical severity.

Considering the discussion above, one could argue that ZTA creates a false sense of security, particularly, when it comes to endpoints, since enterprises that begin ZTA adoption are encouraged to allow access to corporate resources via BYOD, unmanaged or even personal devices, by relying on a mixture of health and security checks and context that can be eventually forged once an endpoint is compromised.

## IV. Potential Solutions to The ZTA Endpoints Problem

As we can see in the previous section, addressing the integrity of the endpoints and detecting compromised endpoints are necessary to improve the effectiveness of ZTAs. In this section, we review some potential approaches and technical solutions to the ZTA endpoints problem.

### A. *Distributed Collaborative Intrusion Detection*

Deploying Intrusion Detection Systems (IDSs) is a well-known approach to effectively detect intrusions based on the anomaly caused by malicious or compromised devices. Hence, it is one of the most promising solutions for the problem discussed in section III. However, implementing a standalone IDS is often insufficient in case of large companies due to the substantial number of false positives and negatives. Shortcomings of standalone IDS systems have been studied by Fung et al. [45], Duma et al. [46] and Weizhi et al. [47]. As a result, DCIDSs have been proposed to improve the efficiency and availability of standalone IDSs.

Collaborative Intrusion Detection Systems (CIDSs) or Collaborative Intrusion Detection Networks (CIDNs) are deployed to eliminate limitations [48] of standalone IDSs. CIDSs consists of cooperating IDSs, using collective knowledge to achieve superior intrusion detection accuracy. Furthermore, DCIDSs deal with various IDS weak cases, such as Distributed Denial of Service (DDoS) attacks. Wu et al. [49] showed that in practice, compared to a standalone IDS setting, CIDSs can reduce the number of missed alarms (to 1 from 7 cases), and they managed to eliminate the number of false alarms in their test system based on Snort, Libsafe, and a new kernel level IDS called Sysmon.

To make this paper as relevant to ZTA in relation to APTs context as possible, we focus our review on three pillars of DCIDSs and the recent advances in the literature for each. Specifically, (1) architecture, (2) alert correlation and (3) alert trustworthiness.

#### 1) *Architecture*

From the architecture perspective, DCIDSs can greatly reduce the rate of false positives and negatives by correlating and analysing multiple suspicious pieces of evidence from diverse sources or sensors throughout the network. There is also potential to decrease computational costs because the intrusion detection resources can be shared between networks. An overview of a DCIDS is shown in Figure 11 [50]. We notice a bidirectional communication in circular format, namely, any detection and correlation unit can potentially connect and communicate with any other unit on the network.

Each participating IDS in the DCIDSs architecture has two core functional units:

- Detection unit, which is responsible for the data collection locally.
- Correlation unit, which is a segment of the overall distributed correlation architecture.

It is worth noting that, despite the benefits brought into the defensive landscape from the DCIDSs, the overall attack surface increases in these architectures, because of their distributed nature. The attackers would have more IDS nodes to target to start working their way towards a stealthy foothold establishment, or simply covering their tracks on a single endpoint. The main security issue identified in the context of DCIDSs is the integrity of the data shared among the IDS nodes, which can be incorrect/incomplete either because of lack of trust (e.g., an IDS node refuses to reveal sensitive data) or the data is sent by a compromised IDS node. Ensuring integrity of the shared data is crucial. Blockchain and the distributed ledger technology can be a promising approach, which we will discuss in the next subsection.

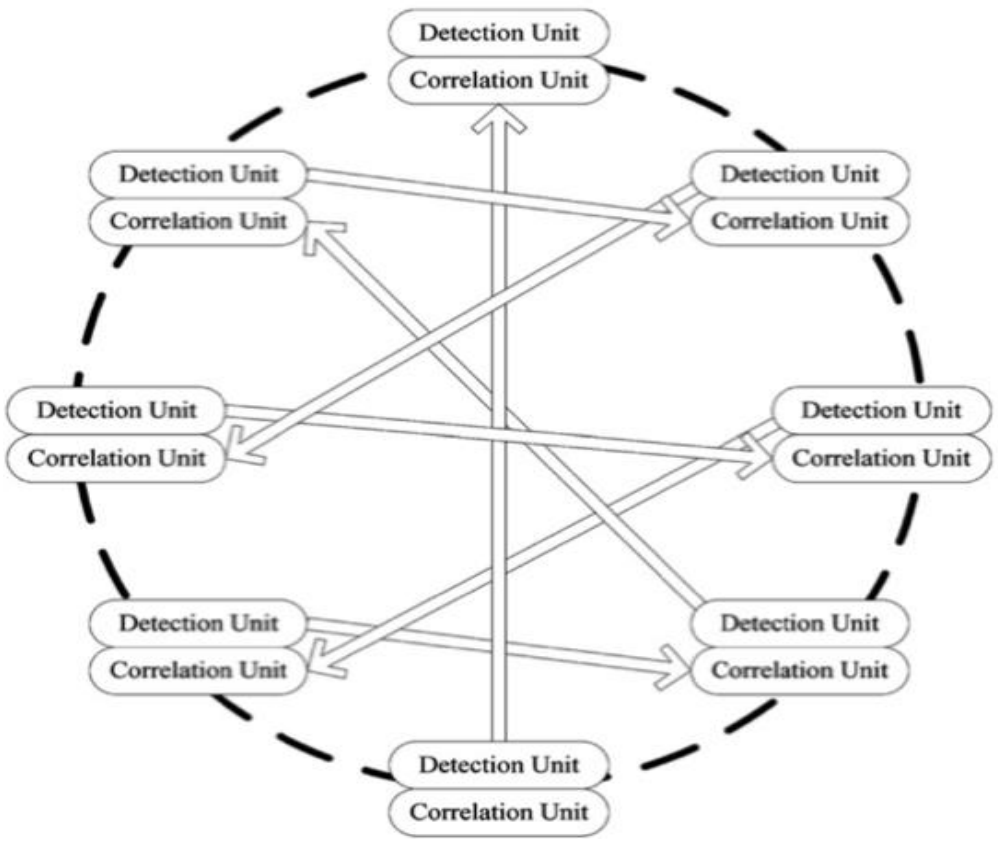

*Figure 11 - DCIDS Reference Architecture [50].*

Another issue in the context of DCIDSs is the dissemination of the alert messages and shared data. Garcia et al. [51] in their study, proposed a DCIDS architecture that correlates alerts from participating nodes effectively via a secure multicast infrastructure, which demonstrated a great capability to detect attacks against and possibly even prevent them. Their architecture was based on local IDS, called “prevention cells”, which detect and record the attack patterns locally. Thereafter, the alert messages were exchanged between the local IDSs to achieve a more effective detection rate.

To cope with APTs, Dash et al. [52] proposed a collaborative host-based IDS approach which detects network intrusion using distributed probabilistic inference. Based on a hierarchical architecture, they proposed three core components in their system: Local Detectors (LDs), being the first component, which serve as a local version of the IDS, analysing the endpoint state and relevant local traffic patterns, secondly, the Global Detectors (GDs) capture the global views of potential attacks by analysing the information gathered through LDs, using a probabilistic model and finally, the Information Sharing System (ISS) which acts as a communication enabler between LDs and GDs via a gossip protocol. In addition, approaches such as binary classifiers are used by LDs to analyse both the incoming and outgoing traffic of the potentially compromised host. Alerts can be triggered if a pre-configured threshold is crossed. The state of the overall security of LDs is constantly transmitted to randomly selected GDs at predefined intervals through the ISS. Finally, the GDs provide global monitoring based on the analysis from data collected from LDs.

This approach could be adapted for the zero-trust context. If an APT had compromised an endpoint within a notional ZTA, or when the attacker had established a foothold on the network, performed data exfiltration from the endpoint, and stolen available credentials, this would be detected. However, detection would be relatively late since the data and credential exfiltration would have already taken place.

*2) Alert Correlation*

In this part, we categorise the DCIDSs based on the alert correlation approaches. These generally include the filter-based approach, the multi-stage approach, the similarity-based approach, and the attack scenario-based approach. In the first case, a prioritisation of alarms takes place based on the criticality of the protected system, while in the second case, the correlation of alerts is based on the causality of former and latter alarms. The third case is simply based on the similarities of alarm attributes. Finally, the attack scenario-based approach is based on predefined attack scenarios.

Dain and Cunningham [53], presented an algorithm that can combine the alerts produced by heterogeneous IDSs via a probabilistic approach. This approach uses three variations of Bayesian Networks (BNs) for effectively detecting network intrusions. Specifically, in the presented algorithm, the CIDS consists of multiple types of IDSs generating alerts, which are converted into an acceptable machine-readable format, and then stored in a standard Structured Query Language (SQL) database. The algorithm then reads the database, categorizes, and relates the alerts into attack scenarios. As soon as new alerts are generated in the IDSs and stored in the database, they are automatically checked against the constructed attack scenario(s).

Cuppens and Ortalo [54] introduced Language to Model a Database for Detection of Attacks (LAMBDA), an attack description language aiming to correlate alerts from various IDSs to CIDSs. LAMBDA can be used to specify the pre and post condition of a target system. Namely, what a system looks like before an attack scenario is launched, and how is it affected after a successful attack scenario. As a result, a wide range of alerts are generated and processed by LAMBDA that eventually are correlated to draw an outcome regarding an ongoing attack scenario or not. During the specification, the overall attack scenario is considered, including all possible threat events and threat types applicable to the target system. In addition, the overall steps for detecting an attack, which might be different in each attack scenario, and the verification of an attack are also considered.

Cheung et al. [55] proposed Correlated Attack Modelling Language (CAML), a modelling language to detect various attack scenarios. Compared to LAMBDA, CAML is also based on the specification of the pre and post condition of the subject system, however, it allows lower-level specification and therefore, lower levels of details are delivered to the IDS nodes. In addition, deep diving into the lower-level specifications provides CAML an advantage when it comes to accurate decision making regarding an ongoing attack.

Templeton and Levitt [56] proposed another attack specification language for DCIDSs, named JIGSAW. Like LAMBDA and CAML, their work is heavily based on pre and post conditions of an attack and the subject target system. A major differentiation with CAML and LAMBDA, however, is that JIGSAW intends to describe specific attacks on the threat event-type level, namely attacks, rather than attack scenarios.

#### 3) *Alert Trustworthiness*

Within a distributed collaborative intrusion detection network, it is imperative to maintain trust between nodes, while also trust the alerts generated by participating nodes. As we mentioned previously, DCIDSs can be particularly effective if IDSs share intrusion-related information with each other; however, the validity and completeness of the information is crucial. In some cases, this is prevented either by compromised devices, or the lack of willingness, as in the case of different organisations to share. Intrusion Detection and Rapid Action (INDRA), a DCIDS approach based on Peer-to-Peer (P2P) infrastructure by Janakiraman et al. [57], proposes an authentication-based solution for alert messages. Specifically, message authentication, based on digital signatures, is used to provide a reasonable level of assurance that alerts are originating from a trusted node by using a central certification authority to authenticate a node's credentials. However, this does not guarantee the completeness and correctness of the messages in the case of compromised nodes or benign nodes that may refuse to 'provide' complete information. Finally, regarding scalability, the central certification authority can be subject to bottleneck as the participating nodes increase.

Chen and Yeager built upon the previous work and proposed the use of "Web of Trust" between participating nodes [58]. The concept is based on the reputation of the nodes, and so the collection, exchange, and evaluation of all information between participants are fully "transparent" to the nodes. Participating nodes can build, over time, a certain level of reputation among themselves, which is ultimately the essence of P2P trust relationships. This approach indeed amplifies the trust bonds required for the purpose of alert broadcasting, in case of an intrusion, and as such it seems promising. However, there is still a problem requiring further study. For example, if a peer takes the necessary time to build a high reputation among the IDS network, then it could potentially broadcast malicious or forged alerts.

### B. *Blockchain Based Intrusion Detection*

Recently, blockchain has been widely investigated as an approach to achieve message integrity in a decentralised or distributed network environment. Blockchain can be either public or private depending on the group of authorised users. Blockchain is closely related to DLT that refers to a database where records of decentralised and transactional data are stored in a sequence (not necessarily grouped in blocks), in a continuous ledger spread through a network and across multiple locations. Blockchain can be considered as a DLT subset, in which batches of transactions are held in blocks, which in turn are linked with hash pointers in a chain [59]. In continuation, each block contains the hash of the previous block in the chain, and therefore, the integrity of each data set in the chain is preserved.

In the following, we review how blockchain has been used to ensure or improve the integrity of shared alert messages and for enforcing trust in IDSs. We start by looking at blockchain types (permissioned vs. permissionless), the consensus mechanisms and finally the related works in the literature for blockchain-enabled IDS. Note that blockchain has been investigated mainly in the context of CIDSs to achieve the integrity of the information shared among the IDSs.

#### 1) *Blockchain Types*

By drawing an analogy between blockchains and databases (as Wüst et al. [60]), we can refer to the blockchain participants as readers and validators (or appenders). A reader refers to a role or entity who can read, analyse, or audit the blockchain. A validator (appender) on the other hand, describes a role or entity that participates in the consensus protocol, collects transactions into a block and finally appends the block to the blockchain. Based on the roles of the participants, we can differentiate between permissionless and permissioned blockchains.

##### a) *Permissionless Blockchains*

In permissionless blockchains, the peers can leave or join the network at any moment, whether they possess the role of a reader or a validator. One of the most interesting parts of this setup is the elimination of a central entity that controls membership overall. Therefore, the written content onto such blockchains is readable by any peer at any given moment. As of today, however, there are implementations using cryptographic primitives that allow for a permissionless blockchain to hide privacy related information. For instance, the Zerocash [61], which acts as a privacy preserving version of Bitcoin. Two prevalent examples of permissionless blockchains include Bitcoin [62] and Ethereum [63].

##### b) *Permissioned Blockchains*

In this setup, a central authority performs the decision making and relevant attribution to peers participating in the read or append roles within the blockchain. Most prevalent examples of permissioned blockchains now are Hyperledger Fabric [64] and R3 Corda [65]. This approach is leaning towards enterprise grade adoption, due to its inherent implementation of a central authority managing peers and their identities. Considering the overly sensitive and confidential use case of blockchain in cyber security and specifically in intrusion detection and prevention, it becomes evident that the permissioned blockchain implementation has better attributes than the permissionless.

It is well-known that blockchains impose computation overhead and extra cost (due to the hash calculations and consensus protocol), and the security of private blockchains greatly depends on the number of the participants. While private blockchains have been implemented by businesses in different sectors such as banks, healthcare, and supply chains[1], mainly to verify the integrity of contracts and secure access to health data, it is still important to see that there are some cases when blockchain is not a suitable solution. Specifically, in our case, we raise the following question: which conditions would make blockchains suitable for the intrusion detection context, and in general cyber security use cases? The "obvious" answer is when multiple entities lack trust in each other, while at the same time wanting to interact with a system and are not willing to agree on a trusted third party. To ease the decision process, Wüst et al. [60] provided a decision flowchart as shown in Figure 12, to help determine whether blockchain addition would be the correct technical solution of a problem. Through a series of simple questions one can conclude if the addition of blockchain would have an added value, and if that is the case, what kind of blockchain would be most suitable (e.g., private, public, permissioned or permissionless).

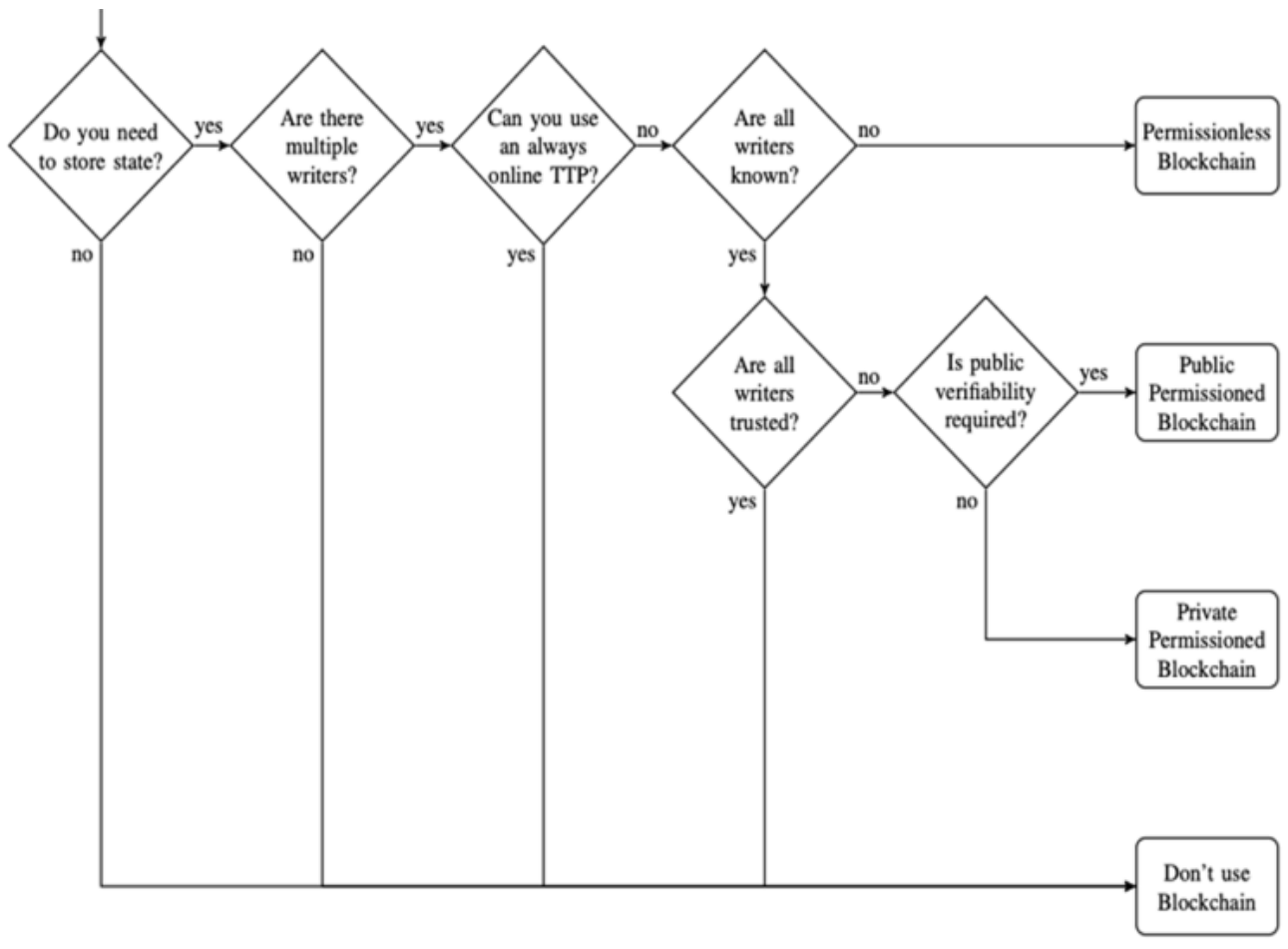

*Figure 12 - Blockchain decision flowchart [60].*

The authors in [60] also provided a performance evaluation among permissioned, permissionless blockchains and a typical database. The results are summarised in Table 4 below, which can help system designers with decision making on blockchain implementations.

In general, blockchain adds complexity, due to the use of consensus mechanisms. Therefore, using a central database or centralised systems enhance the performance in the sense of throughput and latency. On one hand, one can refer to Bitcoin, which is capable of handling 7 transactions per second and can extend up to 66 with no compromise in security. On the other hand, Visa International Service Association Inc. (VISA) an American multinational financial services corporation, which operates a highly centralised system that can manage throughput of approximately fifty thousand transactions. Conclusively, there is a trade-off between scaling and throughput. Specifically, for a blockchain enabled IDS, how well that system would scale to many validators with thousands of hashes as inputs (e.g., detection rules) versus how much throughput such a system would produce in a predefined amount of time. Such trade-offs should be considered when we try to incorporate blockchain elements into intrusion detection.

*Table 4 - Properties of permissionless / permissioned blockchains and central database.*

| | Permissionless blockchain | Permissioned blockchain | Central Database |
|---|---|---|---|
| Throughput | Low | High | Very High |
| Latency | Slow | Medium | Fast |
| Number of readers | High | High | High |
| Number of validators | High | Low | High |
| Number of untrusted users | High | Low | 0 |
| Consensus mechanism | Mainly PoW Some PoS | BFT protocols | None |
| Centrally managed | No | Yes | Yes |

### 2) *The consensus mechanisms*

Assuming a blockchain enabled IDS, where multiple nodes, function as peers, are spread throughout the network for monitoring, gathering and data correlation purposes, they must reach consensus somehow. There must be an effective, practical, dependable, efficient, continuous, and secure mechanism to guarantee that all events and alerts are received and sent and are real and unaltered while all peer members concur to the status of the ledger. That said, there are several consensus mechanisms providing such capabilities, each one with their different attributes [66].

#### a) *Proof of Work (PoW)*

This serves as the most popular consensus protocol and was first introduced in Bitcoin. PoW introduces the roles of the miners, those who are responsible to solve cryptographic puzzles while competing for a reward. However, PoW is probably not suitable for blockchain enabled IDS (within a private enterprise environment) as the concept of rewarded miners would introduce huge security gaps and trust loopholes in the system.

#### b) *Proof of Stake (PoS)*

In this case, there is no competition between the miners. Instead, PoS relies on the validators, who are pseudo-randomly selected to validate a block. In addition, it introduces the so-called stake tokens, where, to participate in this sequence, the validator enrols by staking some of his/her own tokens. Therefore, participants are rewarded based on the number of staked tokens. Considering the blockchain based IDS use case, such a mechanism would create a bottleneck as participants with a high number of tokens staked would automatically have better chances of being selected for validation, which in turn creates a security risk when we talk about events, rules, and alerts of an IDS.

---

[1] Forbes, Blockchain 50, https://www.forbes.com/sites/michaeldelcastillo/2020/02/19/blockchain-50/ (accessed March 2021)

#### c) Practical Byzantine Fault Tolerance (PBFT)

In PBFT, a predefined group of individuals function as validators. Participants must reach consensus when a new event occurs while at the same time, they must verify that no data has been modified during the event transmission. If 2/3 of the participants reach consensus, then the decision is considered final.

#### d) Proof of Burn (PoB) & Proof of Capacity (PoC)

Like the above-mentioned mechanisms PoB and PoC are mining and reward-based mechanisms, which, as outlined above, have an inherent disadvantage when it comes to enterprise grade adoption for the use case of a blockchain enabled IDS, due to confidentiality and integrity reasons [66].

Finally, to summarise this section, a comparative evaluation of the most widely implemented consensus mechanisms can be found in Table 5 (Hazari et al. [66]).

*Table 5 - Consensus mechanisms comparative evaluation.*

| Consensus Mechanisms | PoW | PoS | BFT |
|---|---|---|---|
| Energy Consumption | Requires high amount of energy | Requires less energy consumption | Requires less energy consumption |
| Advanced Hardware Requirement | Required | Not Required | Not Required |
| Centralization | Decentralized | Partially Centralized | Centralized |
| Double Spending Attack | Possible | Difficult | N/A |
| Scalability | Not Scalable | Scalable | Scalable |
| Memory Requirement | Significant due to public ledger | Significant due to public ledger | Less than PoW or PoS |
| Security | Attack with 51% is possible | Attack with 51% not possible | May have a single point of failure |

### 3) Related works on blockchain enabled IDSs

A universal architecture that incorporates CIDS with permissioned blockchain has been described by Alexopoulos et al. [67], together with a design decisions analysis process required when implementing such architecture. In this architecture, a set of intrusion related alerts are defined as transactions within the blockchain. Then, using the consensus protocol, all collaborating IDS nodes can verify the validity of the transactions prior to conveying them into a block. Eventually, the stored set of alerts shall be tamperproof within the blockchain. However, neither implementation details nor practical results are provided in their paper, hence, the idea remains explicitly theoretical.

Similar work at a theoretical level was published by Meng et al. [68], where they studied data and trust management challenges on current IDS architectures. The authors delivered the first review corresponding to the intersection of intrusion detection systems and blockchain technology, while also outlining the prospective application of such collaboration. One of the key conclusions they made was that the blockchain technology can greatly assist in enhancing an intrusion detection system's core tasks such as trust computation, exchange of alerts and data sharing.

A step further in detecting adversaries via blockchain enabled cyber defence capabilities was addressed by Li et al. [69]. They specifically studied the integrity property in CIDS, by considering a highly likely scenario which we often encounter nowadays, namely, insider attacks such as a malicious node generating forged signatures and then sharing it throughout peers. If that scenario becomes a reality, intruders could potentially remain undetected, which would greatly affect the effectiveness of a CIDS. In addition, the authors used the blockchain technology to solve the subject issue in a verifiable manner and evaluated the results via a so-called Collaborative Blockchained Signature-Based Intrusion Detection System (CBSigIDS) development, a generic framework of CIDS based on blockchain. Figure 13 below depicts a high-level overview of the proposed blockchain based CIDS framework.

On the other hand, a more practical approach was proposed by Golomb et al. [70], namely, a Collaborative IoT Anomaly Detection (CIoTA) framework. This is a lightweight framework that leverages blockchain technology to accomplish collaborative and distributed anomaly detection. In this framework, blockchain is being used to incrementally feed an anomaly detection model and establish consensus among IoT devices. Eventually, the authors created their own distributed IoT simulation platform consisting of 48 Raspberry Pi's to evaluate and demonstrate CIoTA's ability to enhance security via blockchain.

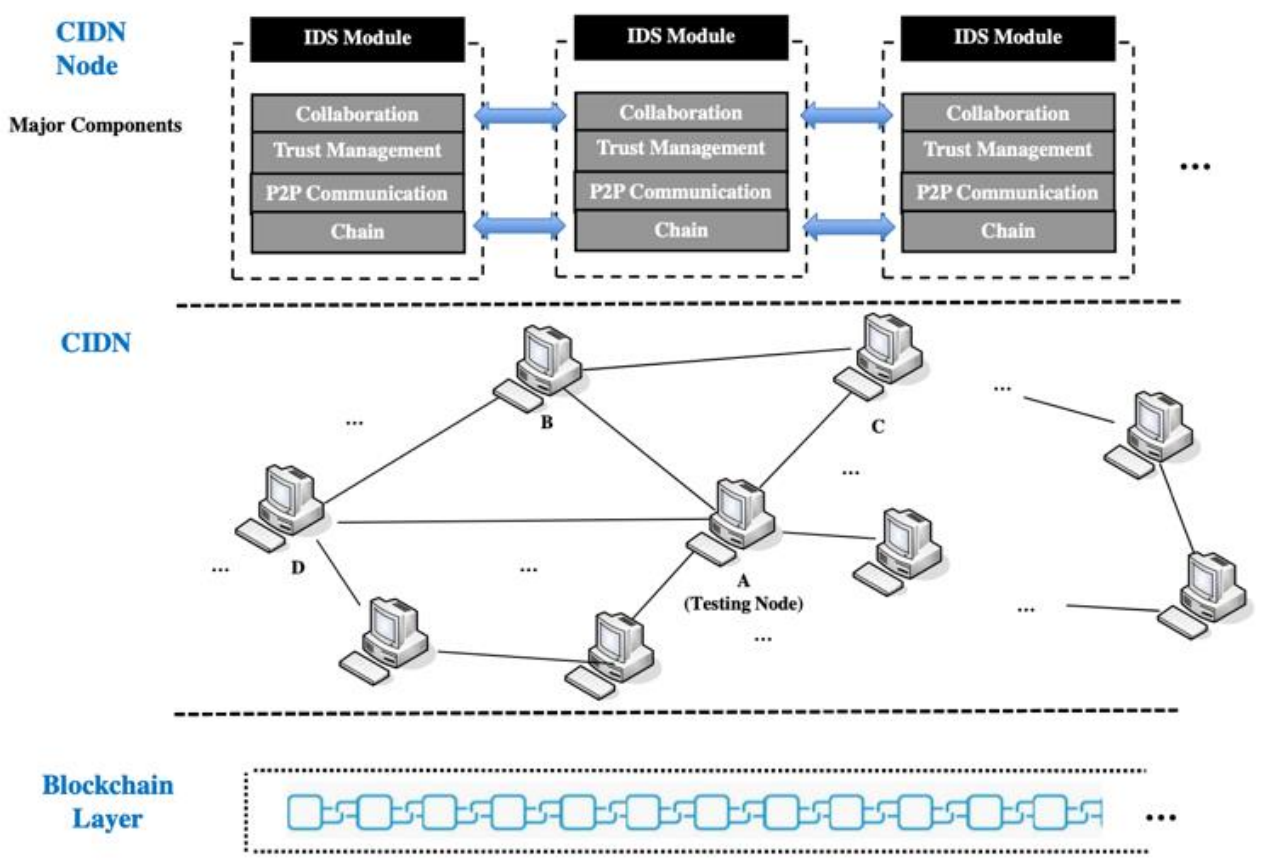

*Figure 13 - High level overview of blockchain based CIDN [66].*

Conclusively, we can say that the previous works validate, mainly at the theoretical level, the potential of blockchain enhancing intrusion detection. There is, however, a practical demonstration of the above conclusion performed by Golomb et al. [70] with CIoTA, although its focus and scope are limited to IoT. Moreover, an IoT network is different from an enterprise network in the sense that it provides less control maturity compared to the current applicable control frameworks and standards. Besides the immense potential of using blockchains in intrusion detection (and prevention), there are probably other advantages that require further research. For instance, a blockchain enabled IDS can be a trusted source of logging, which in turn can further enhance and maximise trust in auditing.

One of the core principles of ZTA, namely, "never trust but verify", seems to match greatly with blockchains' inherent attribution where every transaction must be validated, consensus must always be achieved, while ledger's immutability seals integrity.

### C. *ZTA and Distributed Blockchain-based IDS Convergence*

In this section, we build upon the ZTA core principle of assuming breach (see section II.C) to discuss how blockchain-based IDS can be employed. For this discussion, we use an example of a ZTA enabled notional bank network, where we assume that a single endpoint has been compromised via a spear phishing attack. As per our review (see section II), and the abovementioned assumption, the lateral movement is highly unlikely once ZTA is in full force [4], adhering to all principles and all mandated controls in place. However, the endpoint itself remains compromised, together with the already authenticated and authorised sessions of the subject user in the endpoint. Moreover, the adversaries can abuse the authenticated and authorised sessions of the user and extend their attack to the systems in reach of the subject user.

DCIDSs, as reviewed in subsection A, would be able to detect such attacks via a plethora of methodologies. Specifically, the attack scenario-based approach for alert correlation when used by DCIDS is an effective and efficient approach for adversary detection. A major shortcoming can be identified, however, with this approach. In the context of ZTA and APTs, (1) the adversaries characteristically use legitimate tools in a malicious manner, and (2) they also use advanced evasive techniques against the standard controls (e.g., signature based / heuristic-based anti-virus etc.) Therefore, the attack scenarios can fluctuate greatly. Until the attack scenario-based approach eventually constructs the relevant and matching scenario, adversaries probably have already established a stealthy foothold into the network, deeming the detection process ineffective, again, in a ZTA context. In addition, the integrity of DCIDSs nodes is questionable as per the literature review in certain scenarios. Our assumption of an APT compromising an endpoint is subject to the same scenario since a determined adversary would likely try to influence the integrity of a node and/or tamper with logs and audit trails to render the attack invisible.

Blockchain based intrusion detection as reviewed in sub-section B, greatly increases the integrity of the audit trail and log files, as well as the overall integrity of the information stored in the blocks themselves. Additionally, blockchain could potentially enhance the efficiency of intrusion detection by extending the immutability aspect of the context of each single identity. Specifically, zero trust security health checks can be used to create the so-called endpoint context. This context, then, could be further fortified by the distributed ledger technology to achieve integrity. ZTA, DCIDSs and blockchain technology seem to have a great intersection and many potential use cases. In fact, some use cases could even be extended beyond detection, to implement blockchain based prevention capabilities.

## V. Summary And Discussion

### A. *Challenges to the Integration of Blockchain and ZTA*

As we can see, ZTA and blockchain take a different approach on trust management, security, and architecture overall, in contrast to the traditional, perimeter-based approach. Table 6 shows the previously mentioned intersection elements in ZTA and blockchain, in contradiction to the traditional perimeter-based approach.

*Table 6 - ZTA & blockchain intersection elements.*

| | **Traditional Perimeter-Based Architecture** | **Zero Trust Architecture** | **Blockchain** |
|---|---|---|---|
| **Overall Approach** | Centralised | Decentralised | Decentralised |
| **Architectural focus** | Perimeter-Focused | Borderless / Distributed | Distributed |
| **Infrastructure trust level** | Trusted or semi-trusted in some cases | Untrusted or trust but verify in some cases | Untrusted |

In perimeter-based approaches, we have the element of centralisation, and the architectural focus is to protect the perimeter. This means that trusted data and assets are placed behind an extremely strict perimeter, assuming that anyone and anything inside that perimeter is trusted, either partially or fully, to access those resources. Ultimately, maximum effort is put into making sure that adversaries will not be able to get beyond that perimeter, while at the same time authorised and authenticated users can still access the data and resources behind it.

This is vastly different from ZTA and blockchain based technologies, which both run in a borderless and decentralised manner. Since there is no perimeter on both ZTA and blockchain, security comes from efficient and effective management of trust. In fact, for blockchain, security comes from the incredible amount of repetition because every node is being asked to keep the same copy of the ledger and periodically reach majority consensus on what the proper data in that ledger should be. As such, the amount of work that an attacker would have to do is practically impossible if adversaries wanted to change, hack, or alter the ledger. That said, it seems that blockchain and ZTA can complement each other in various use cases, since both share at least some fundamental principles.

Determined attackers, such as in case of APTs, with the necessary knowledge and resources have demonstrated their ability to compromise various endpoints with ease, and plant malware to establish footholds into corporate networks. The different ZTA deployment models (see section II.E) and implementations (see section II.F) are great instruments in the hands of defenders, in their effort to prevent lateral movement. The result is a highly secure, trustless and borderless architecture with fine grained identity-based access controls always seeking to verify. However, the endpoints are still the Achilles heel of ZTA. Adversaries can potentially tamper with ZTA's security health checks once an endpoint is compromised, therefore leveraging the already authenticated and authorised user's session.

### B. Future Directions

Blockchain technology can enhance ZTA implementations in several use cases. As described in section IV.C, a blockchain-based intrusion detection system could help in amplifying the detection capability. At the same time, it is possible to fortify the backend storage of relevant logs and audit trails in the blockchain, providing immutability. Blockchain-based authentication could also be used to enhance remote working. For instance, a blockchain based layer could be added on top of an SDP to strengthen the endpoint's integrity. Enhancing the prevention capability with blockchain is of equal, if not more, interest. Combining a blockchain-based intrusion detection and prevention system would ultimately augment ZTA onto the endpoints, significantly enhancing the detection and prevention capabilities.

However, issues such as performance, computing overhead and choosing the right implementation of blockchain remain the main questions to adopting this approach. These questions need further research to answer sufficiently.

## VI. Conclusion

In this paper, we provided a state-of-the-art review on zero-trust and ZTAs, which are relevant and emerging research and development areas. Based on 43 papers in the ZTA literature, we reviewed several aspects of the zero trust approaches and open questions. We discussed the main differences between traditional perimeter-based models and zero trust approaches. In addition, the core tenets and core capabilities of the zero-trust concept were presented, with different existing theoretical and real-world implementations of ZTAs. Thereafter, based on examples, we discussed the potential security problems with current ZTAs, and outlined some potential and promising approaches that can be used to tackle those problems. Specifically, one of the approaches we explored is the possibility of adapting DLT and blockchain to verify the integrity of the endpoints in a ZTA. Based on the state-of-the-art in this area, we concluded that DLTs and blockchain can play a critical part in augmenting one of the core tenets of zero trust architectures, namely, the assumed breach mindset. However, their implementation requires thoughtful consideration due to computation overhead and the trade-off between security and usability.

## VII. References

[1] R. Rapuzzi and M. Repetto, "Building situational awareness for network threats in fog/edge computing: Emerging paradigms beyond the security perimeter model," *Future Generation Computer Systems,* vol. 85, pp. 235-249, August 2018.

[2] E. GIlman and D. Barth, Zero Trust Networks: Building Secure Systems in Untrusted Networks 1st Edition, A. Courtney and V. Wilson, Eds., O'Reilly, 2017, pp. 21-29,51-62,65-90,93-101,113-125,137-171,173-207,209-215.

[3] J. Forum™, "The Open Group," May 2007. [Online]. Available: https://collaboration.opengroup.org/jericho/commandments_v1.2.pdf. [Accessed October 2020].

[4] N. S. A. (NSA), "U.S. Department of Defense," 25 February 2021. [Online]. Available: https://media.defense.gov/2021/Feb/25/2002588479/-1/1/0/CSI_EMBRACING_ZT_SECURITY_MODEL_20UOO115131-21.PDF [Accessed February 2021].

[5] R. Ward and B. Beyer, "BeyondCorp - A new approach to enterprise security," *BeyondCorp,* vol. 39, no. 6, pp. 6-11, 2014.

[6] D. Teixeira, A. Singh and M. Agarwal, "Evade Antiviruses, bypass firewalls and exploit complex environments with the most widely used penetration testing framework," in *Metasploit Penetration Testing Cookbook, Third Edition*, Birmingham - Mumbai, Packt Publishing Ltd., 2018, pp. 264-269, 188-229.

[7] S. Rose, O. Borchert, S. Mitchell and S. Connelly, "nist.gov," 20 August 2020. [Online]. Available: https://doi.org/10.6028/NIST.SP.800-207. [Accessed 17 10 2020].

[8] M. Campbell, "Beyond Zero Trust: Trust Is a Vulnerability," *Computer Society,* vol. 53, no. 10, pp. 110-113, 2020.

[9] M. Pincheira, M. Vecchio, R. Giaffreda and S. S. Kanhere, "Exploiting constrained IoT devices in a trustless blockchain-based water management system," in *020 IEEE International Conference on Blockchain and Cryptocurrency (ICBC)*, Toronto, Canada, 2020.

[10] P. Kasireddy, "medium.com," 3 February 2018. [Online]. Available: https://medium.com/@preethikasireddy/eli5-what-do-we-mean-by-blockchains-are-trustless-aa420635d5f6. [Accessed 17 October 2020].

[11] P. J. Taylor, T. Dargahi, A. Dehghantanha, R. M. Parizi and K.-K. Raymond Choo, "A systematic literature review of blockchain cyber security," *Digital Communications and Networks,* vol. 6, no. 2, pp. 147-156, May 2020.

[12] M. Tayyab, B. Belaton and M. Anbar, "ICMPv6-Based DoS and DDoS Attacks Detection Using Machine Learning Techniques, Open Challenges, and Blockchain Applicability: A Review," *IEEE Access,* vol. 8, no. 170529, pp. 13-14,16-17, 28 September 2020.

[13] M. Zhou, L. Han, H. Lu and C. Fu, "Distributed collaborative intrusion detection system for vehicular Ad Hoc networks based on invariant," *Computer Networks,* vol. 172, no. 107174, pp. 12-14, 8 May 2020.

[14] W. Li, Y. Wang, Z. Jin, K. Yu, J. Li and Y. Xiang, "g, Challenge-based Collaborative Intrusion Detection in Software Defined Networking: An Evaluation," *Digital Communications and Networks,* vol. 10, no. 1016, pp. 4-6, 19 September 2020.

[15] J. Kindervag, S. Balaouras and L. Coit, "Build Security Into Your Network's DNA: The Zero Trust Network Architecture," Forrester Research, Inc., Cambridge, MA 02139 USA, 2010.

[16] J. G. Grimes, "acqnotes.com," June 2007. [Online]. Available: http://www.acqnotes.com/Attachments/DoD%20GIG%20Architectural%20Vision,%20June%2007.pdf. [Accessed October 2020].

[17] R. Ward and B. Beyer, "BeyondCorp: A New Approach to Enterprise Security," *Usenix,* vol. 39, no. 6, pp. 6-10, December 2014.

[18] L. Cittadini, B. Spear, B. Beyer and M. Saltonstall, "BeyondCorp: The Access Proxy," *Security,* vol. 41, no. 4, pp. 28-33, 2016.

[19] B. Osborn, J. McWilliams, B. Beyer and M. Saltonstall, "BeyondCorp: Design to Deployment at Google," *Security,* vol. 41, no. 1, pp. 28-34, 2016.

[20] J. Peck, B. Beyer, C. Beske and M. Saltonstall, "Migrating to BeyondCorp: Maintaining Productivity While Improving Security.," *Security,* vol. 42, no. 2, pp. 49-55, 2017.

[21] V. Escobedo, B. Beyer, M. Saltonstall and F. Żyźniewski, "BeyondCorp 5: The User Experience," *Security,* vol. 42, no. 3, pp. 38-43, 2017.

[22] H. King, M. Janosko, B. (. E. Beyer and M. Saltonstall, "BeyondCorp 6: Building a Healthy Fleet.," *Security,* vol. 43, no. 3, pp. 24-30, 2018.

[23] C. Smith, "Understanding concepts in the defence in depth strategy," in *IEEE 37th Annual 2003 International Carnahan Conference on Security Technology, 2003. Proceedings.*, Taipei, Taiwan, 2003.

[24] D. Pallais, "Microsoft," Microsoft, 18 September 2019. [Online]. Available: https://www.microsoft.com/en-us/microsoft-365/blog/2019/09/18/why-banks-adopt-modern-cybersecurity-zero-trust-model/#:~:text=Many%20banks%20today%20still%20rely,protect%20data%20from%20malicious%20attacks.&text=So%2C%20whether%20an%20insider%20acts,data%2. [Accessed 23 October 2020].

[25] C. Cunningham, "forrester.com," Forrester Research, Inc., 27 March 2018. [Online]. Available: https://go.forrester.com/blogs/next-generation-access-and-zero-trust/. [Accessed October 2020].

[26] C. DeCusatis, P. Liengtiraphan, A. Sager and M. Pinelli, "Implementing Zero Trust Cloud Networks with Transport Access Control and First Packet Authentication," in *2016 IEEE*

*International Conference on Smart Cloud (SmartCloud)*, New York, NY, USA, November 2016.

[27] M. Samaniego and R. Deters, “Zero-Trust Hierarchical Management in IoT.,” in *2018 IEEE International Congress on Internet of Things (ICIOT)*, San Francisco, CA, USA, 2018.

[28] Marketsandmarkets, “Zero-Trust Security Market by Solution Type (Data Security, Endpoint Security, API Security, Security Analytics, Security Policy Management), Deployment Type, Authentication Type, Organization Size, Vertical, and Region - Global Forecast to 2024,” Marketsandmarkets, 2019.

[29] B. Embrey, “The top three factors driving zero trust adoption,” *Computer Fraud & Security*, vol. 2020, no. 9, pp. 13-15, 22 September 2020.

[30] S. Mehraj and T. M. Banday, “Establishing a Zero Trust Strategy in Cloud Computing Environment,” in *2020 International Conference on Computer Communication and Informatics (ICCCI)*, Coimbatore, India, 2020.

[31] Y. Xiangshuai and W. Huijuan, “Survey on Zero-Trust Network Security,” in *Artificial Intelligence and Security. ICAIS 2020. Communications in Computer and Information Science.*, Singapore, 2020.

[32] S. Keeriyattil, “Microsegmentation and Zero Trust: Introduction.,” in *Zero Trust Networks with VMware NSX.*, Berkeley, CA, Apress, 2019.

[33] R. Mital, “IMPROVING TRUST IN A ZERO TRUST ARCHITECTURE (ZTA),” *Getting it right - Collaborating for mission success*, vol. 10, no. 4, p. 2, June 2020.

[34] J. Koilpillai and N. A. Murray, “Cloud Security Alliance,” CSA, 5 May 2020. [Online]. Available: https://cloudsecurityalliance.org/software-defined-perimeter/. [Accessed October 2020].

[35] Gartner, “Gartner Research,” 4 March 2020. [Online]. Available: https://www.gartner.com/teamsiteanalytics/servePDF?g=/imagesrv/media-products/pdf/Qi-An-Xin/Qi-An-Xin-1-1OKONUN2.pdf. [Accessed October 2020].

[36] C. Tankard, “Advanced Persistent threats and how to monitor and deter them.,” *Network Security*, vol. 2011, no. 8, pp. 16-19, August 2011.

[37] M. Labs, “Wired,” 3 March 2010. [Online]. Available: https://www.wired.com/images_blogs/threatlevel/2010/03/operationaurora_wp_0310_fnl.pdf. [Accessed 26 October 2020].

[38] Google, “BeyondProd: A new approach to cloud-native security,” 2020.

[39] M. Barcelo, A. Correa, J. Llorca, A. M. Tulino, L. J. Vicario and A. Morell, “IoT-Cloud Service Optimization in Next Generation Smart Environments,” *EEE Journal on Selected Areas in Communications*, vol. 32, no. 12, pp. 4077-4090, 25 December 2016.

[40] B. Anggorojati, P. N. Mahalle, R. N. Prasad and R. Prasad, “Capability-based access control delegation model on the federated IoT network,” in *the 15th International Symposium on Wireless Personal Multimedia Communications*, Taipei, 2012.

[41] S. Keeriyattil, Zero Trust Networks with VMware NSX, Berkeley, CA: Apress, 2019, pp. 173-177.

[42] E. M. Hutchins, M. J. Cloppert and R. M. Amin, “Lockheed Martin Corporation,” Lockheed Martin Corporation, 5 May 2015. [Online]. Available: https://www.lockheedmartin.com/content/dam/lockheed-martin/rms/documents/cyber/LM-White-Paper-Intel-Driven-Defense.pdf. [Accessed October 2020].

[43] NetMarketShare, “NetMarketShare.com,” NetApplications.com, 17 October 2020. [Online]. Available: https://netmarketshare.com/. [Accessed 17 October 2020].

[44] MITRE, “The MITRE Corporation,” MITRE, 18 October 2020. [Online]. Available: https://mitre-attack.github.io/attack-navigator/enterprise/. [Accessed 18 October 2020].

[45] C. J. Fung, O. Baysal, Z. Jie, I. Aib and R. Boutaba, “Trust Management for Host-Based Collaborative Intrusion Detection,” in *DSOM 2008: Managing Large-Scale Service Deployment*, Berlin, Heidelberg, 2008.

[46] C. Duma, M. Karresand, N. Shahmehri and G. Caronni, “A Trust-Aware, P2P-Based Overlay for Intrusion Detection.,” in *17th International Workshop on Database and Expert Systems Applications (DEXA'06)*, Krakow, Poland, 2006.

[47] M. Weizhi, L. Wenjuan and K. Lam-For, “Design of intelligent KNN-based alarm filter using knowledge-based alert verification in intrusion detection,” in *Security and Communication Networks 8(18)*, 2015.

[48] A. Khraisat, I. Gondal, P. Vamplew and J. Kamruzzaman, “Survey of intrusion detection systems: techniques, datasets and challenges,” *Cybersecurity*, vol. 20, no. 2, pp. 50-62, 17 July 2019.

[49] Y.-S. Wu, B. Foo, Y. Mei and S. Bagchi, “Collaborative Intrusion Detection System (CIDS): A Framework for Accurate and Efficient IDS,” in *Computer Security Applications Conference, 2003. Proceedings. 19th Annual*, 2004.

[50] V. Z. Chenfeng, C. Leckie and S. Karunasekera, “A survey of coordinated attacks and collaborative intrusion detection,” *Computers & Security*, no. 29, pp. 124-140, 29 June 2009.

[51] J. Garcia, F. Autrel, J. Borrell, S. Castillo, F. Cuppens and G. Navarro, “Decentralized publish-subscribe system to prevent coordinated attacks via alert correlation.,” in *Sixth international conference on information and communications security*, Berlin, Heidelber, 2004.

[52] D. Dash, B. Kveton, J. M. Agosta, E. Schooler, J. Chandrashekar, A. Bachrach and A. Newman, “When Gossip is Good: Distributed Probabilistic Inference for Detection of Slow Network Intrusions.,” in T*he Twenty-First National Conference on Artificial Intelligence and the Eighteenth Innovative Applications of Artificial Intelligence Conference.*, Boston, Massachusetts, USA, 2006.

[53] O. Dain and R. K. Cunningham, “Fusing A Heterogeneous Alert Stream into Scenarios.,” in *Applications of Data Mining in Computer Security.*, vol. 6, Boston, MA., Springer, 2002.

[54] F. Cuppens and R. Ortalo, “LAMBDA: A Language to Model a Database for Detection of Attacks.,” in *International Workshop on Recent Advances in Intrusion Detection.*, Berlin, Heidelberg, 2000.

[55] S. Cheung, U. Lindqvist and M. Fong, “Modeling multistep cyber-attacks for scenario recognition.,” in *Proceedings DARPA Information Survivability Conference and Exposition.*, Washington, DC, USA, USA, 2003.

[56] S. J. Templeton and K. Levitt, “A requires/provides model for computer attacks.,” in *Proceedings of new security paradigms workshop.*, 2001.

[57] R. Janakiraman, M. Waldvoger and Q. Zhang, “Indra: a peer-to-peer approach to network intrusion detection and prevention,” in *WET ICE 2003. Proceedings. Twelfth IEEE International Workshops on Enabling Technologies: Infrastructure for Collaborative Enterprises, 2003.*, Linz, Austria, Austria, 2003.

[58] R. Chen and W. Yeager, “Poblano A Distributed Trust Model for Peer-to-Peer Networks.,” IEEE, 2001.

[59] G. Verdian, P. Tasca, C. Paterson and G. Mondelli, “Quant Network,” 31 January 2018. [Online]. Available: https://www.quant.network/wp-content/uploads/2020/07/Quant_Overledger_Whitepaper-Sep-1.pdf. [Accessed 17 November 2020].

[60] K. Wüst and A. Gervais, “IACR,” 2017. [Online]. Available: https://eprint.iacr.org/2017/375.pdf. [Accessed March 2021].

[61] E. Ben-Sasson, A. Chiesa, C. Garman, M. Green, I. Miers, E. Tromer and M. Virza, “Zerocash: Decentralized Anonymous Payments from Bitcoin,” in *IEEE Security & Privacy Symposium*, 2014.

[62] S. Nakamoto, “bitcoin.org,” 2009. [Online]. Available: https://bitcoin.org/bitcoin.pdf. [Accessed 22 March 2021].

[63] V. Buterin, “ethereum.org,” 19 March 2021. [Online]. Available: https://ethereum.org/en/whitepaper/. [Accessed March 2021].

[64] Hyperledger, “Hyperledger,” March 2020. [Online]. Available: https://www.hyperledger.org/wp-content/uploads/2020/03/hyperledger_fabric_whitepaper.pdf. [Accessed 22 March 2021].

[65] R3, “R3.com,” August 2019. [Online]. Available: https://www.r3.com/reports/corda-technical-whitepaper/. [Accessed 22 March 2021].

[66] S. S. Hazari and Q. H. Mahmoud, “Comparative evaluation of consensus mechanisms cryptocurrencies,” WILEY, 2019.

[67] N. Alexopoulos, E. Vasilomanolakis, N. R. Ivánkó and M. Mühlhäuser, “Towards Blockchain-Based Collaborative Intrusion Detection Systems,” in *International Conference on Critical Information Infrastructures Security*, 2018.

[68] W. Meng, E. Wolfgang Tischhauser, Q. Wang, Y. Wang and J. Han, “When Intrusion Detection Meets Blockchain Technology: A

Review," *IEEE Access,* vol. 6, no. 1, pp. 10179-10188, 15 March 2018.

[69] W. Li, S. Tug, W. Meng and Y. Wang, "Designing collaborative blockchained signature-based intrusion detection in IoT environments," *Future Generation Computer Systems,* vol. 96, pp. 481-489, July 2019.

[70] T. Golomb, Y. Mirsky and Y. Elovici, "CIoTA: Collaborative IoT Anomaly Detection via Blockchain," in *Proceedings of workshop on Decentralized IoT Security and Standards (DISS),* Negev, 2018.

[71] R. A. Cormier, N. T. Spurgeon, D. L. Schuh, P. A. Smyton, R. S. Swarz, F. C. Wendt and G. Rebovich Jr, mitre.org, Bedford, MA: MITRE Corporate Communications and Public Affairs, 2014, pp. 167-174.

[72] M. Coole, J. Corkill and A. Woodward, "Defence in Depth, Protection in Depth and Security in Depth: A Comparative Analysis Towards a Common Usage Language," in *Proceedings of the 5th Australian Security and Intelligence Conference,* Perth, Western Australia, 2012.

[73] NetMarketShare, "NetMarketShare.com," NetApplications.com, 17 October 2020. [Online]. Available: https://netmarketshare.com/operating-system-market-share.aspx?options=%7B%22filter%22%3A%7B%22%24and%22%3A%5B%7B%22deviceType%22%3A%7B%22%24in%22%3A%5B%22Desktop%2Flaptop%22%5D%7D%7D%5D%7D%2C%22dateLabel%22%3A%22Trend%22%2C%22attributes%22%3A%22share%22%2. [Accessed 17 October 2020].

[74] M. Steichen, S. Hommes and R. State, "ChainGuard — A firewall for blockchain applications using SDN with OpenFlow," in *Principles, Systems and Applications of IP Telecommunications (IPTComm),* Chicago, 2017.

[75] L. Zhichum, Y. Chen and A. Beach, "Towards Scalable and Robust Distributed Intrusion Alert Fusion with Good Load Balancing.," in *Proceedings of the 2006 SIGCOMM Workshop on Large-Scale Attack Defense (LSAD),* 2006.

[76] T. A. Tuan, "A Game-Theoretic Analysis of Trust Management in P2P Systems," in *2006 First International Conference on Communications and Electronics,* Hanoi, Vietnam, 2006.

[77] L. Wenjuan, M. Yuxin and K. Lam-For, "Enhancing Trust Evaluation Using Intrusion Sensitivity in Collaborative Intrusion Detection Networks: Feasibility and Challenges," in *2013 Ninth International Conference on Computational Intelligence and Security,* Leshan, 2013.

[78] L. Wenjuan, M. Weizhi and K. Lam-For, "Design of Intrusion Sensitivity-Based Trust Management Model for Collaborative Intrusion Detection Networks," in *Proceedings of the 8th IFIP WG 11.11 International Conference on Trust Management (IFIPTM),* Berlin, Heidelberg, 2014.

[79] K. Uttecht, "Zero Trust (ZT) Concepts for Federal Government Architectures," Department of Homeland Security (DHS) Science and Technology Directorate (S&T), Lexington, Massachusetts, 2020.

[80] F. Valeur, G. Vigna, C. kruegel and R. A. Kemmerer, "Comprehensive approach to intrusion detection alert correlation," *Transactions on Dependable and Secure Computing.,* vol. 1, no. 3, pp. 2-8, July-September 2004.

[81] P. Ning, Y. Cui and D. S. Reeves, "Constructing attack scenarios through correlation of intrusion alerts," in *Proceedings of the 9th ACM conference on Computer and communications security,* 2002.